\documentclass{svmult}
\usepackage{bm}
\usepackage{graphicx}
\usepackage{amsmath}

\DeclareMathOperator{\Img}{\mathrm{Im}}
\DeclareMathOperator{\Rea}{\mathrm{Re}}
\DeclareMathOperator{\Sp}{\mathrm{Sp}}
\begin{document}

\title{Crossed Andreev reflection and spin-resolved non-local electron transport}
\titlerunning{Crossed Andreev reflection} 
\author{Mikhail S. Kalenkov$^1$ \and Andrei D. Zaikin$^{2,1}$}
\institute{I.E. Tamm Department of Theoretical Physics, P.N.
Lebedev Physical Institute, 119991 Moscow, Russia \and Institut f\"ur Nanotechnologie, Karlsruher Institut f\"ur Tehnologie (KIT), 76021, Karlsruhe, Germany}

\maketitle

\begin{abstract}
The phenomenon of crossed
Andreev reflection (CAR) is known to play a key role in
non-local electron transport across
three-terminal normal-superconducting-normal (NSN)
devices. Here we review our general theory of non-local charge
transport in three-terminal disordered ferromagnet-superconductor-ferromagnet (FSF)
structures. We demonstrate that CAR is highly sensitive to electron
spins and yields a rich variety of properties of non-local
conductance which we describe non-perturbatively at arbitrary
voltages, temperature, degree of disorder, spin-dependent interface
transmissions and their polarizations. We demonstrate that magnetic effects
have different implications: While strong exchange field suppresses disorder-induced
electron interference in ferromagnetic electrodes, spin-sensitive electron scattering
at SF interfaces can drive the total non-local conductance negative at sufficiently
low energies. At higher energies magnetic effects become less important and the
non-local resistance behaves similarly to the non-magnetic case.
Our results can be applied to multi-terminal
hybrid structures with normal, ferromagnetic and half-metallic
electrodes and can be directly tested in future experiments.
\end{abstract}

\section{Introduction}
In hybrid NS structures quasiparticle current flowing in a normal
metal is converted into that of Cooper pairs inside a
superconductor. For quasiparticle energies above the
superconducting gap $\varepsilon > \Delta$ this conversion is
accompanied by electron-hole imbalance which
relaxes deep inside a superconductor. At subgap energies
$\varepsilon < \Delta$ the physical picture is entirely different.
In this case
quasiparticle-to-Cooper-pair current conversion is provided by the
mechanism of Andreev reflection (AR) \cite{And}: A quasiparticle enters
the superconductor from the normal metal at a length of order of
the superconducting coherence length $\xi_S$, forms a Cooper pair
together with another quasiparticle, while a hole goes back into
the normal metal. As a result, the net charge $2e$ is transferred through
the NS interface which acquires non-zero subgap conductance
down to $T=0$ \cite{BTK}.

\begin{figure}
\centerline{\includegraphics[width=75mm]{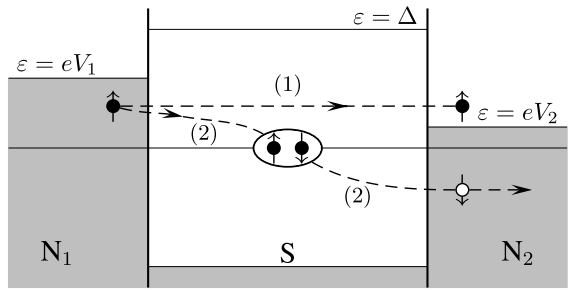}}
\caption{Two elementary processes contributing to non-local
conductance of an NSN device: (1) direct electron transfer and (2)
crossed Andreev reflection.}
\label{car-fig}
\end{figure}

AR remains essentially
a local effect provided there exists only one NS interface in the
system or, else, if the distance between different NS interfaces
greatly exceeds the superconducting coherence length $\xi$. If,
however, the distance $L$ between two adjacent NS interfaces (i.e.
the superconductor size) is smaller than (or comparable with)
$\xi$, two additional {\it non-local} processes come into play
(see Fig.~\ref{car-fig}). One such process corresponds to direct electron
transfer between two N-metals through a superconductor. Another
process is the so-called crossed Andreev reflection \cite{BF,GF}
(CAR): An electron penetrating into the superconductor from the
first N-terminal may form a Cooper pair together with another
electron from the second N-terminal. In this case a hole will go
into the second N-metal and AR becomes a non-local effect.
This phenomenon of CAR enables
direct experimental demonstration of entanglement between
electrons in spatially separated N-electrodes and can strongly
influence non-local transport of electrons in hybrid NSN systems.

Non-local electron transport in the presence of CAR was recently
investigated both experimentally
\cite{Beckmann,Teun,Venkat,Venkat2,Basel,Deutscher,Basel2,Beckmann2} and theoretically
\cite{FFH,KZ06,BG,Belzig,Duhot,GZ07,LY,KZ07,KZ07E,KZ08,GZ09,GKZ,Bergeret,GZ10}
(see also further references therein)
demonstrating a rich variety of physical processes involved in the
problem. It was shown \cite{FFH} that in the lowest order in the interface
transmission and at $T=0$ CAR contribution to cross-terminal
conductance is exactly canceled by that from elastic electron
cotunneling (EC), i.e. the non-local conductance vanishes in this
limit. Taking into account higher order processes in barrier transmissions
eliminates this feature and yields non-zero values of
cross-conductance \cite{KZ06}.

Another interesting issue is the effect of disorder. It is well
known that disorder enhances interference effects and, hence, can
strongly modify local subgap conductance of NS interfaces in the
low energy limit \cite{VZK,HN,Z}. Non-local conductance of
multi-terminal hybrid NSN structures in the presence of disorder
was studied in a number of papers \cite{BG,Belzig,Duhot,GZ07,GKZ,Bergeret}.
A general quasiclassical theory was constructed by Golubev and the present
authors \cite{GKZ}. It was demonstrated that an interplay
between CAR, quantum interference of electrons and non-local
charge imbalance dominates the behavior of diffusive NSN systems
being essential for quantitative interpretation of a number of
experimental observations \cite{Venkat,Basel,Deutscher}.
In particular, strong enhancement of non-local spectral
conductance was predicted at low energies due to quantum interference
of electrons in disordered N-terminals. At the same time, non-local
resistance $R_{12}$ remains smooth at small
energies and, furthermore, was found to depend neither on
parameters of NS interfaces nor on those of N-terminals. At higher
temperatures $R_{12}$ was shown to exhibit a peak caused by the trade-off
between charge imbalance and Andreev reflection.

Yet another interesting subject is an interplay between CAR and
Coulomb interaction. The effect of electron-electron interactions
on AR was investigated in a number of papers \cite{Z,HHK,GZ06}.
Interactions should also affect CAR, e.g., by lifting the exact
cancellation of EC and CAR contributions \cite{LY} already in the
lowest order in tunneling. A similar effect can occur in the presence
of external ac fields \cite{GZ09}. A general theory of
non-local transport in NSN systems with disorder and electron-electron
interactions was very recently developed by Golubev and one of the present authors
\cite{GZ10} direct relation between Coulomb effects and non-local shot noise.
In the tunneling limit non-local differential conductance is found to have
an S-like shape and can turn negative at non-zero bias. At high transmissions
CAR turned out to be responsible both for positive noise cross-correlations and
for Coulomb anti-blockade of non-local electron transport.

An important property of both AR and CAR is that these processes
should be sensitive to magnetic properties of normal electrodes
because these processes essentially depend on spins of scattered
electrons. One possible way to demonstrate spin-resolved CAR is to
use ferromagnets (F) instead of normal electrodes
ferromagnet-superconductor-ferromagnet (FSF) structures.
First experiments on such FSF
structures \cite{Beckmann} illustrated this point by demonstrating
the dependence of non-local conductance on the polarization of
ferromagnetic terminals. Hence, for better understanding of
non-local effects in multi-terminal hybrid proximity structures it
is necessary to construct a theory of {\it spin-resolved}
non-local transport. In the case of ballistic systems in the
lowest order order in tunneling this
task was accomplished in \cite{FFH}. Here we will
generalize our quasiclassical
approach \cite{KZ06,GKZ} to explicitly focus on spin
effects and construct a theory of non-local electron
transport in both ballistic and diffusive NSN and FSF
structures with spin-active interfaces beyond lowest order
perturbation theory in their transmissions.

The structure of the paper is as follows. In Sec. 2 we will
describe non-local spin-resolved electron transport in
ballistic NSN structures with spin-active interfaces. In Sec. 3 we
will further extend our formalism and evaluate both local and
non-local conductances in SFS structures in the presence of disorder.
Our main conclusions are briefly summarized in Sec. 4.

\section{Spin-resolved transport in ballistic systems}

\begin{figure}
\centerline{\includegraphics[width=65mm]{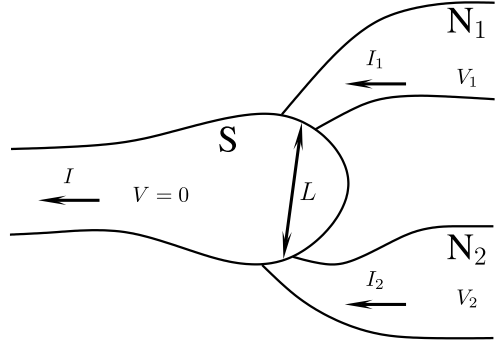}}
\caption{Schematics of our NSN device.} \label{nsn}
\end{figure}

Let us consider three-terminal NSN structure depicted in Fig.~\ref{nsn}. We will assume that all three metallic electrodes are
non-magnetic and ballistic, i.e. the electron elastic mean free
path in each metal is larger than any other relevant size scale.
In order to resolve spin-dependent effects we will assume that
both NS interfaces are spin-active, i.e. we will distinguish
``spin-up'' and ``spin-down'' transmissions of the first
($D_{1\uparrow}$ and $D_{1\downarrow}$) and the second
($D_{2\uparrow}$ and $D_{2\downarrow}$) SN interface. All these
four transmissions may take any value from zero to one. We also
introduce the angle $\varphi$ between polarizations of two
interfaces which can take any value between 0 and $2\pi$.

In what follows effective cross-sections of the two interfaces
will be denoted respectively as $\mathcal{A}_1$ and
$\mathcal{A}_2$. The distance between these interfaces $L$ as well
as other geometric parameters are assumed to be much larger than
$\sqrt{\mathcal{A}_{1,2}}$, i.e. effectively both contacts are
metallic constrictions. In this case the voltage drops only across
SN interfaces and not inside large metallic electrodes.

For convenience, we will set the electric potential of the
S-electrode equal to zero, $V=0$. In the presence of bias voltages
$V_1$ and $V_2$ applied to two normal electrodes (see Fig.~\ref{nsn}) the currents $I_1$ and $I_2$ will flow through SN$_1$
and SN$_2$ interfaces. These currents can be evaluated with the
aid of the quasiclassical formalism of nonequilibrium
Green-Eilenberger-Keldysh functions \cite{BWBSZ} $\hat g^{R,A,K}$
which we briefly specify below.

\subsection{Quasiclassical equations}

In the ballistic limit the corresponding Eilenberger equations
take the form
\begin{gather}
\begin{split}
\left[
\varepsilon \hat\tau_3+
eV(\bm{r},t)-
\hat\Delta(\bm{r},t),
\hat g^{R,A,K} (\bm{p}_F, \varepsilon, \bm{r},t)
\right]
\\+
i\bm{v}_F \nabla \hat g^{R,A,K} (\bm{p}_F, \varepsilon, \bm{r},t) =0,
\end{split}
\label{Eil}
\end{gather}
where $[\hat a, \hat b]= \hat a\hat b - \hat b \hat
a$, $\varepsilon$ is the quasiparticle energy, $\bm{p}_F=m\bm{v}_F$ is the
electron Fermi momentum vector and $\hat\tau_3$ is the Pauli matrix in Nambu
space.
The functions  $\hat g^{R,A,K}$ also obey the normalization conditions
$(\hat g^R)^2=(\hat g^A)^2=1$ and $\hat g^R \hat g^K + \hat g^K \hat g^A =0$.
Here and below the product of matrices is defined as time convolution.

Green functions $\hat g^{R,A,K}$ and $\hat\Delta$ are $4\times4$ matrices in
Nambu and spin spaces. In Nambu space they can be parameterized as
\begin{equation}
        \hat g^{R,A,K} =
        \begin{pmatrix}
                g^{R,A,K} & f^{R,A,K} \\
                \tilde f^{R,A,K} & \tilde g^{R,A,K} \\
        \end{pmatrix}, \quad
        \hat\Delta=
        \begin{pmatrix}
                0 & \Delta i\sigma_2 \\
                \Delta^* i\sigma_2& 0 \\
        \end{pmatrix},
\end{equation}
where $g^{R,A,K}$, $f^{R,A,K}$, $\tilde f^{R,A,K}$, $\tilde g^{R,A,K}$ are
$2\times2$ matrices in the spin space, $\Delta$ is the BCS order parameter and
$\sigma_i$ are Pauli matrices. For simplicity we will only consider the case of
spin-singlet isotropic pairing in the superconducting electrode.
The current density is
related to the Keldysh function $\hat g^K$ according to the standard relation
\begin{equation}
\bm{j}(\bm{r}, t)= \dfrac{e N_0}{8} \int d \varepsilon
\left< \bm{v}_F \mathrm{Sp} [\hat \tau_3 \hat g^K(\bm{p}_F,
\varepsilon, \bm{r},t)] \right>,
\label{current_bal}
\end{equation}
where $N_0=mp_F/2\pi^2$ is the density of state at the Fermi level and
angular brackets $\left< ... \right>$ denote averaging over the Fermi momentum.

\subsection{Riccati parameterization}

The above matrix Green-Keldysh functions can be conveniently
parameterized by four Riccati amplitudes $\gamma^{R,A}$, $\tilde
\gamma^{R,A}$ and two ``distribution functions'' $x^K$, $\tilde
x^K$ (here and below we chose to follow the notations
\cite{Eschrig00}):
\begin{equation}
\hat g^K=
2
\hat N^R
\begin{pmatrix}
x^K - \gamma^R  \tilde x^K  \tilde \gamma^A &
-\gamma^R  \tilde x^K + x^K  \gamma^A \\
-\tilde \gamma^R  x^K + \tilde x^K  \tilde \gamma^A &
\tilde x^K - \tilde \gamma^R  x^K  \gamma^A \\
\end{pmatrix}
\hat N^A,
\label{gkparam}
\end{equation}
where functions $\gamma^{R,A}$ and $\tilde \gamma^{R,A}$ are Riccati amplitudes
\begin{equation}
\hat g^{R,A}=\pm
    \hat N^{R,A}
    \begin{pmatrix}
    1+\gamma^{R,A} \tilde \gamma^{R,A} & 2\gamma^{R,A} \\
    -2 \tilde \gamma^{R,A} & -1- \tilde \gamma^{R,A}  \gamma^{R,A} \\
    \end{pmatrix}
    \label{graparam}
\end{equation}
and $\hat N^{R,A}$ are the following matrices
\begin{equation}
\hat N^{R,A}=
    \begin{pmatrix}
    (1-\gamma^{R,A} \tilde \gamma^{R,A})^{-1} & 0 \\
    0 & (1-\tilde \gamma^{R,A}  \gamma^{R,A} )^{-1} \\
    \end{pmatrix}.
    \label{nrparam}
\end{equation}
With the aid of the above parameterization one can identically transform
the quasiclassical equations \eqref{Eil} into the following set of
effectively decoupled equations for
Riccati amplitudes and distribution functions \cite{Eschrig00}
\begin{gather}
\begin{split}
i\bm{v}_F \nabla \gamma^{R,A}+ [\varepsilon+eV(\bm{r},t)]\gamma^{R,A}+
\gamma^{R,A}[\varepsilon-eV(\bm{r},t)]
\\
=\gamma^{R,A}\Delta^* i\sigma_2\gamma^{R,A}-\Delta i\sigma_2,
\end{split}
\label{eqgamma}
\\
\begin{split}
i\bm{v}_F \nabla \tilde\gamma^{R,A}- [\varepsilon-eV(\bm{r},t)]\tilde\gamma^{R,A}-
\tilde\gamma^{R,A}[\varepsilon+eV(\bm{r},t)]
\\
=\tilde\gamma^{R,A}\Delta i\sigma_2\tilde\gamma^{R,A}-\Delta^* i\sigma_2,
\end{split}
\label{eqtildegamma}
\\
\begin{split}
i\bm{v}_F \nabla
x^K+[\varepsilon+eV(\bm{r},t)]x^K-x^K[\varepsilon+eV(\bm{r},t)]
\\
-\gamma^{R}\Delta^* i\sigma_2 x^K-
x^K \Delta i\sigma_2 \tilde\gamma^{A}=0,
\end{split}
\label{eqx}
\\
\begin{split}
i\bm{v}_F \nabla \tilde x^K-
[\varepsilon-eV(\bm{r},t)]\tilde x^K+\tilde x^K[\varepsilon-eV(\bm{r},t)]
\\
-\tilde \gamma^{R}\Delta i\sigma_2 \tilde x^K-
\tilde x^K \Delta^* i\sigma_2 \gamma^{A}=0.
\end{split}
\label{eqtildex}
\end{gather}

Depending on the particular trajectory it is also convenient to
introduce a ``replica'' of both Riccati amplitudes and
distribution functions which -- again following the notations
\cite{Eschrig00,Zhao04} -- will be denoted by capital letters
$\Gamma$ and $X$. These ``capital'' Riccati amplitudes and
distribution functions obey the same equations
\eqref{eqgamma}-\eqref{eqtildex} with the replacement $\gamma
\rightarrow \Gamma$ and $x \rightarrow X$. The distinction between
different Riccati amplitudes and distribution functions will be
made explicit below.

\subsection{Boundary conditions}

\begin{figure}
\centerline{\includegraphics[width=75mm]{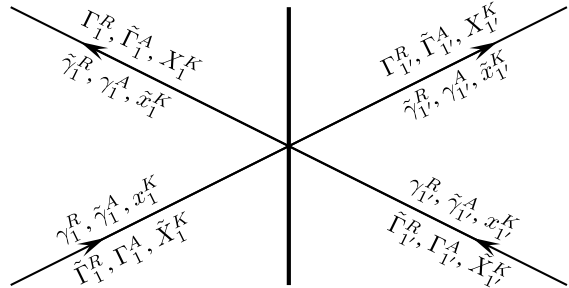}} \caption{Riccati
amplitudes for incoming and outgoing trajectories from the both
sides of the interface. } \label{sis}
\end{figure}

Quasiclassical equations should be supplemented by appropriate
boundary conditions at metallic interfaces. In the case of
specularly reflecting spin-degenerate interfaces these conditions
were derived by Zaitsev \cite{Zaitsev} and later generalized to
spin-active interfaces \cite{Millis88}, see also \cite{Eschrig09}
for recent review on this subject.

Before specifying these conditions it is important to emphasize
that the applicability of the Eilenberger quasiclassical formalism
with appropriate boundary conditions to hybrid structures with two
or more barriers is, in general, a non-trivial issue
\cite{GZ02,OS}. Electrons scattered at different barriers
interfere and form bound states (resonances) which cannot be
correctly described within the quasiclassical formalism employing
Zaitsev boundary conditions or their direct generalization. Here
we avoid this problem by choosing the appropriate geometry of our
NSN device, see Fig.~\ref{nsn}. In our system any relevant trajectory
reaches each NS interface only once whereas the probability of
multiple reflections at both interfaces is small in the parameter
$\mathcal{A}_1 \mathcal{A}_2/L^4 \ll 1$. Hence, resonances formed
by multiply reflected electron waves can be neglected, and our
formalism remains adequate for the problem in question.

It will be convenient for us to formulate the boundary conditions
directly in terms of Riccati amplitudes and the distribution
functions. Let us consider the first NS interface and explicitly
specify the relations between Riccati amplitudes and distribution
functions for incoming and outgoing trajectories, see
Fig.~\ref{sis}. The boundary conditions for $\Gamma_1^R$,
$\tilde\Gamma_1^A$ and $X_1^K$ can be written in the form
\cite{Zhao04}
\begin{gather}
\Gamma_1^R= r_{1l}^R\gamma_1^R \underline{S}_{11}^+ +
t_{1l}^R\gamma_{1'}^R {\underline{S}}_{11'}^+,
\\
\tilde\Gamma_1^A= \underline{S}_{11} \tilde\gamma_1^A \tilde r_{1r}^A  +
{\underline{S}}_{11'} \tilde\gamma_{1'}^A \tilde t_{1r}^A ,
\\
X_1^K=r_{1l}^R x_1^K \tilde r_{1r}^A +
t_{1l}^R x_{1'}^K \tilde t_{1r}^A -
a_{1l}^R \tilde x_{1'}^K \tilde a_{1r}^A.
\end{gather}
Here we defined the transmission ($t$), reflection  ($r$), and
branch-conversion ($a$) amplitudes as:
\begin{gather}
r_{1l}^R=+[(\beta_{1'1}^R)^{-1}S_{11}^+ - (\beta_{1'1'}^R)^{-1}S_{11'}^+]^{-1}
(\beta_{1'1}^R)^{-1},
\\
t_{1l}^R=-[(\beta_{1'1}^R)^{-1}S_{11}^+ - (\beta_{1'1'}^R)^{-1}S_{11'}^+]^{-1}
(\beta_{1'1'}^R)^{-1},
\\
\tilde r_{1r}^A=+(\beta_{1'1}^A)^{-1}
[S_{11}(\beta_{1'1}^A)^{-1} - S_{11'}(\beta_{1'1'}^A)^{-1}]^{-1},
\\
\tilde t_{1r}^A=-(\beta_{1'1'}^A)^{-1}
[S_{11}(\beta_{1'1}^A)^{-1} - S_{11'}(\beta_{1'1'}^A)^{-1}]^{-1},
\\
a_{1l}^R=(\Gamma_1^R \underline{S}_{11} - S_{11}\gamma_1^R)(\tilde
\beta_{11'}^R)^{-1},
\\
\tilde a_{1r}^A=(\tilde \beta_{11'}^A)^{-1}
(\underline{S}_{11}^+ \tilde \Gamma_1^A  - \tilde \gamma_1^A S_{11}^+),
\end{gather}
where
\begin{gather}
\beta_{ij}^R=S_{ij}^+ - \gamma_j^R \underline{S}_{ij}^+ \tilde\gamma_i^R,\
\tilde\beta_{ij}^R=\underline{S}_{ji} - \tilde \gamma_j^R S_{ji}\gamma_i^R,
\\
\beta_{ij}^A=S_{ij} - \gamma_i^A \underline{S}_{ij} \tilde\gamma_j^A,\
\tilde\beta_{ij}^A=\underline{S}_{ji}^+ - \tilde \gamma_i^A S_{ji}^+\gamma_j^A.
\end{gather}
Similarly, the boundary conditions for $\tilde\Gamma_1^R$,
$\Gamma_1^A$, and $\tilde X_1^K$ take the form:
\begin{gather}
\tilde\Gamma_1^R= \tilde r_{1l}^R\tilde\gamma_1^R S_{11} +
\tilde t_{1l}^R\tilde\gamma_{1'}^R S_{1'1},
\\
\Gamma_1^A= S_{11}^+ \gamma_1^A r_{1r}^A  +
S_{1'1}^+ \gamma_{1'}^A t_{1r}^A ,
\\
\tilde X_1^K=\tilde r_{1l}^R \tilde x_1^K r_{1r}^A +
\tilde t_{1l}^R \tilde x_{1'}^K t_{1r}^A -
\tilde a_{1l}^R x_{1'}^K a_{1r}^A,
\end{gather}
where
\begin{gather}
\tilde r_{1l}^R=+[(\tilde \beta_{1'1}^R)^{-1}\underline{S}_{11} -
(\tilde \beta_{1'1'}^R)^{-1}\underline{S}_{1'1}]^{-1}
(\tilde\beta_{1'1}^R)^{-1},
\\
t_{1l}^R=-[(\tilde \beta_{1'1}^R)^{-1}\underline{S}_{11} -
(\tilde \beta_{1'1'}^R)^{-1}\underline{S}_{1'1}]^{-1}
(\tilde\beta_{1'1'}^R)^{-1},
\\
r_{1r}^A=+(\tilde\beta_{1'1}^A)^{-1}
[\underline{S}_{11}^+(\tilde\beta_{1'1}^A)^{-1} -
\underline{S}_{1'1}^+(\tilde\beta_{1'1'}^A)^{-1}]^{-1},
\\
\tilde t_{1r}^A=-(\tilde\beta_{1'1'}^A)^{-1}
[\underline{S}_{11}^+(\tilde\beta_{1'1}^A)^{-1} -
\underline{S}_{1'1}^+(\tilde\beta_{1'1'}^A)^{-1}]^{-1},
\\
\tilde a_{1l}^R=(\tilde \Gamma_1^R S_{11}^+ - \underline{S}_{11}^+ \tilde \gamma_1^R)( \beta_{11'}^R)^{-1},
\\
a_{1r}^A=(\beta_{11'}^A)^{-1} (S_{11} \Gamma_1^A  - \gamma_1^A \underline{S}_{11}).
\end{gather}
Boundary conditions for $\Gamma_{1'}^{R,A}$, $\tilde
\Gamma_{1'}^{R,A}$, $X^K_{1'}$ and $\tilde X^K_{1'}$ can be
obtained from the above equations simply by replacing $1
\leftrightarrow 1'$.

The matrices $S_{11}$, $S_{11'}$, $S_{1'1}$, and $S_{1'1'}$
constitute the components of the $\mathcal{S}$-matrix describing
electron scattering at the first interface:
\begin{equation}
\mathcal{S}=
\begin{pmatrix}
S_{11} & S_{11'}\\
S_{1'1} & S_{1'1'}\\
\end{pmatrix}, \quad
\mathcal{S}\mathcal{S}^+=1
\end{equation}

\begin{figure*}
\centerline{
\includegraphics{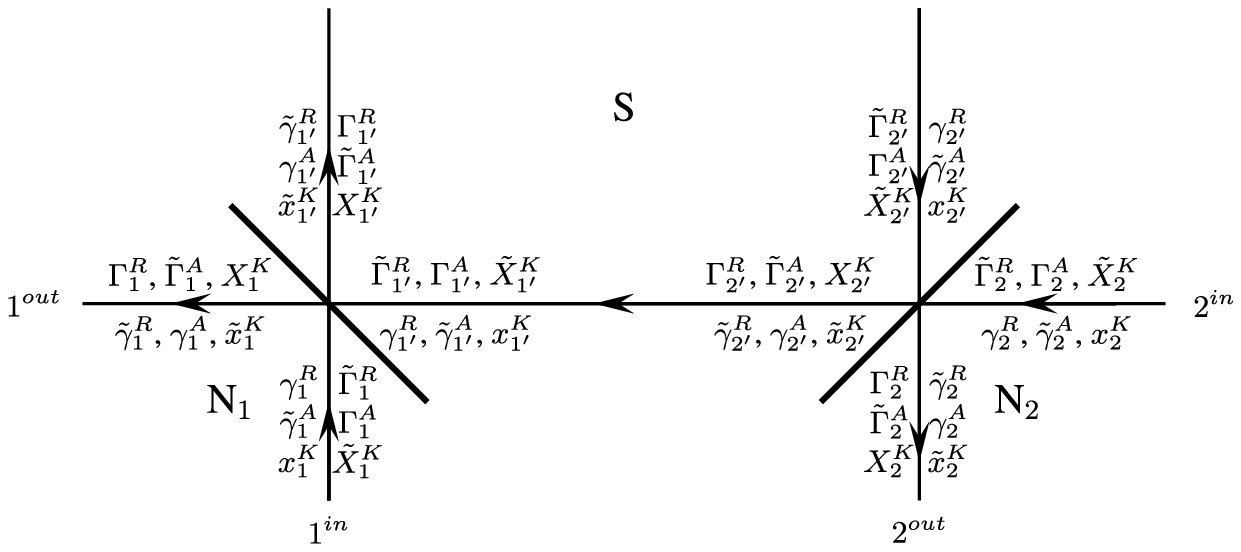}
} \caption{Riccati amplitudes for incoming and outgoing
trajectories for an NSN structure with two barriers. The arrows
define quasiparticle momentum directions. We also indicate
relevant Riccati amplitudes and distribution functions
parameterizing the Green-Keldysh function for the corresponding
trajectory.} \label{traject_b}
\end{figure*}

In our three terminal geometry nonlocal conductance arises only
from trajectories that cross both interfaces, as illustrated in
Fig.~\ref{traject_b}. Accordingly, the above boundary conditions
should be employed at both NS interfaces.

Finally, one needs to specify the asymptotic boundary conditions
far from NS interfaces. Deep in metallic electrodes we have
\begin{gather}
\gamma_1^R=\tilde \gamma_1^R=\gamma_1^A=\tilde \gamma_1^A=0,
\\
x_1^K=h_0(\varepsilon+eV_1),\quad
\tilde x_1^K=-h_0(\varepsilon-eV_1),
\\
\gamma_2^R=\tilde \gamma_2^R=\gamma_2^A=\tilde \gamma_2^A=0,
\\
x_2^K=h_0(\varepsilon+eV_2),\quad
\tilde x_2^K=-h_0(\varepsilon-eV_2),
\end{gather}
where $h_0(\varepsilon)=\tanh(\varepsilon/2T)$ - equilibrium distribution function.
In the bulk of superconducting electrode we have
\begin{gather}
\tilde \gamma_{1'}^R=-a(\varepsilon)i\sigma_2, \quad
\gamma_{1'}^A=a^*(\varepsilon)i\sigma_2,
\\
\tilde x_{1'}^K=-[1-|a(\varepsilon)|^2]h_0(\varepsilon),
\\
\gamma_{2'}^R=a(\varepsilon)i\sigma_2, \quad
\tilde\gamma_{2'}^A=-a^*(\varepsilon)i\sigma_2,
\\
x_{2'}^K=[1-|a(\varepsilon)|^2]h_0(\varepsilon),
\end{gather}
where we denoted $a( \varepsilon ) = - (\varepsilon - \sqrt{
\varepsilon^2 - \Delta^2})/\Delta$.

\subsection{Green functions}

With the aid of the above equations and boundary conditions it is
straightforward to evaluate the quasiclassical Green-Keldysh
functions for our three-terminal device along any trajectory of
interest. For instance, from the boundary conditions at the second
interface we find
\begin{equation}
\Gamma_{2'}^R=ia(\varepsilon)A_2\sigma_2,
\label{GGG}
\end{equation}
where $A_2=S_{2'2'}\sigma_2\underline{S}_{2'2'}^+ \sigma_2$.
Integrating Eq. \eqref{eqgamma} along the trajectory connecting
both interfaces and using Eq.~(\ref{GGG}) as the initial condition
we immediately evaluate the Riccati amplitude at the first
interface:
\begin{gather}
\gamma_{1'}^R=
i\dfrac{aA_2 + (aA_2\varepsilon + \Delta)Q
}{1-(aA_2\Delta + \varepsilon)Q}\sigma_2,
\\
Q=\dfrac{\tanh\left[i\Omega L/v_F\right]}{
\Omega}, \quad
\Omega = \sqrt{\varepsilon^2-\Delta^2}.
\end{gather}
Employing the boundary conditions again we obtain
\begin{gather}
\Gamma_1^R=iS_{11'}K_{21}^{-1}
\left[aA_2+(aA_2\varepsilon + \Delta)Q\right]
\sigma_2\underline{S}_{11'}^+,
\\
\tilde\Gamma_1^R=-i a \underline{S}_{1'1}^+ \sigma_2 S_{1'1'}
K_{21}^{-1}\left[1-(aA_2\Delta+\varepsilon)Q\right]
S_{1'1'}^{-1} S_{1'1},
\end{gather}
where
\begin{gather}
K_{ij}= (1-a^2 A_i A_j) - \left[ \varepsilon (1+a^2 A_i A_j) + \Delta a (A_i+A_j) \right]Q,
\\
A_1=\sigma_2 \underline{S}_{1'1'}^+\sigma_2 S_{1'1'}.
\end{gather}
We also note that the relation $(\Gamma^{R,A})^+ = \tilde \Gamma^{A,R}$ and $(\gamma^{R,A})^+ = \tilde \gamma^{A,R}$
makes it unnecessary (while redundant) to separately
calculate the advanced Riccati amplitudes.

Let us now evaluate the distribution functions at both interfaces.
With the aid of the boundary conditions at the second interface we
obtain
\begin{multline}
X_{2'}^K=
S_{2'2'}S_{2'2'}^+\left(1-|a|^2\right)h_0(\varepsilon) +
S_{2'2}S_{2'2}^+ x_2^K
\\-
|a|^2 S_{2'2'} \sigma_2 \underline{S}_{22'}^+\underline{S}_{22'} \sigma_2
S_{2'2'}^+ \tilde x_2^K.
\end{multline}
Integrating Eq. \eqref{eqx} along the trajectory connecting both
interfaces with initial condition for $X_{2'}^K$, we arrive at the
expression for $x_{1'}^K$
\begin{multline}
x_{1'}^K=
\left[1-(a A_2\Delta + \varepsilon)Q\right]^{-1} X_{2'}^K
\\\times
(1-\tanh^2iL\Omega/v_F)
\left[1-(aA_2\Delta + \varepsilon)Q\right]^{+^{-1}}.
\end{multline}
Then we can find distribution functions at the first interface. On
the normal metal side of the interface we find
\begin{equation}
X_1^K=
r_{1l}^R x_1^K r_{1l}^{R^+} +
t_{1l}^R x_{1'}^K t_{1l}^{R^+} +
a_{1l}^R  a_{1l}^{R^+} \left(1-|a|^2\right)h_0(\varepsilon)
\end{equation}
where
\begin{gather}
\begin{split}
r_{1l}^R=S_{11'}K_{21}^{-1}
\Bigl[ (1-(aA_2\Delta + \varepsilon)Q)S_{1'1'}^+S_{1'1}^{+^{-1}}
\\-
a(aA_2+(aA_2\varepsilon + \Delta)Q) \sigma_2 \underline{S}_{1'1'}^+ \sigma_2 S_{1'1}^{+^{-1}} \Bigr],
\end{split}
\\
t_{1l}^R=S_{11'}K_{21}^{-1} (1-(aA_2\Delta + \varepsilon)Q),
\\
a_{1l}^R=i S_{11'}K_{21}^{-1}(aA_2+(aA_2\varepsilon + \Delta)Q)\sigma_2 \underline{S}_{1'1'}^+.
\end{gather}
The corresponding expression for $\tilde X_1^K$ is obtained
analogously. We get
\begin{equation}
\tilde X_1^K= \tilde r_{1l}^R \tilde x_1^K \tilde r_{1l}^{R^+} -
\tilde t_{1l}^R \tilde t_{1l}^{R^+} \left(1-|a|^2\right)h_0(\varepsilon) -
\tilde a_{1l}^R x_{1'}^K \tilde a_{1l}^{R^+}.
\end{equation}
where
\begin{gather}
\begin{split}
\tilde r_{1l}^R=-
\Bigl[ \underline{S}_{1'1}^{-1}\underline{S}_{1'1'} \sigma_2 (1-(aA_2\Delta + \varepsilon)Q)
\\-
\underline{S}_{1'1}^{-1} \sigma_2 S_{1'1'} a(aA_2+(aA_2\varepsilon + \Delta)Q)  \Bigr]
K_{12}^{-1} \sigma_2 \underline{S}_{11'}^+,
\end{split}
\\
\tilde t_{1l}^R=\underline{S}_{1'1}^+ \underline{S}_{1'1'}^{+^{-1}} \sigma_2
(1-(aA_2\Delta + \varepsilon)Q) K_{12}^{-1}
\sigma_2 \underline{S}_{1'1'}^+,
\\
\tilde a_{1l}^R= i a \underline{S}_{1'1}^+ \sigma_2 S_{1'1'}
 K_{21}^{-1} (1-(aA_2\Delta + \varepsilon)Q).
\end{gather}
Combining the above results for the Riccati amplitudes and the
distribution functions we can easily evaluate the Keldysh Green
function at the first interface. For instance, for the trajectory
$1^{out}$ (see Fig.~\ref{traject_b}) we obtain
\begin{equation}
g^K_{1^{out}}=2(X_1^K - \Gamma_1^R \tilde x_1^K \Gamma_1^{R^+}), \quad
\tilde g^K_{1^{out}}=2\tilde x_1^K.
\end{equation}
The Keldysh Green function for the trajectory $1^{in}$ is
evaluated analogously, and we get
\begin{equation}
g^K_{1^{in}}=2 x_1^K, \quad
\tilde g^K_{1^{in}}=2(\tilde X_1^K-\tilde\Gamma_1^R x_1^K \tilde\Gamma_1^{R^+}).
\end{equation}

\subsection{Nonlocal conductance: General results}

Now we are ready to evaluate the current $I_1$ across the first
interface. This current takes the form:
\begin{equation}
I_1=I_1^{BTK}(V_1)-\dfrac{G_0}{8e} \int d \varepsilon \Sp
(\hat\tau_3 \hat g^K_{1^{out}} - \hat\tau_3 \hat g^K_{1^{in}}),
\label{current_g}
\end{equation}
where
\begin{equation}
G_0=\frac{8\gamma_1 \gamma_2
\mathcal{N}_1\mathcal{N}_2}{R_qp_F^2L^2}
\end{equation}
is the normal state conductance of our device at fully transparent
interfaces, $p_F\gamma _{1(2)}$ is normal to the first (second)
interface component of the Fermi momentum for electrons
propagating straight between the interfaces,
$\mathcal{N}_{1,2}=p_F^2\mathcal{A}_{1,2}/4\pi$ define the number
of conducting channels of the corresponding interface,
$R_q=2\pi/e^2$ is the quantum resistance unit.

Here $I_1^{BTK}(V_1)$ stands for the contribution to the current
through the first interface coming from trajectories that never
cross the second interface. This is just the standard BTK
contribution \cite{BTK,Zhao04}. The non-trivial contribution is
represented by the last term in Eq. (\ref{current_g}) which
accounts for the presence of the second NS interface. We observe
that this non-local contribution to the current is small as
$\propto 1p_F^2L^2$. This term will be analyzed in details
below.

The functions $\hat g^K_{1^{in}}$ and $\hat g^K_{1^{out}}$ are the
Keldysh Green functions evaluated on the trajectories $1^{in}$ and
$1^{out}$ respectively. Using the above expression for the Riccati
amplitudes and the distribution functions we find
\begin{multline}
\Sp (\hat\tau_3 \hat g^K_{1^{out}} - \hat\tau_3 \hat g^K_{1^{in}})
= 2 \Sp[r_{1l}^R r_{1l}^{R^+} - \tilde \Gamma_1^R  \tilde
\Gamma_1^{R^+} -1] (h_0(\varepsilon+eV_1)-h_0(\varepsilon))
\\-
2
\Sp[\tilde r_{1l}^R \tilde r_{1l}^{R^+}- \Gamma_1^R \Gamma_1^{R^+}
-1] (h_0(\varepsilon-eV_1)-h_0(\varepsilon))
\\+
2
(1-\tanh^2iL\Omega/v_F)  \Sp[K_{21}^{-1}
\{S_{2'2}S_{2'2}^+ (h_0(\varepsilon+eV_2)-h_0(\varepsilon))
\\+
|a|^2 S_{2'2'} \sigma_2 \underline{S}_{22'}^+\underline{S}_{22'}
\sigma_2 S_{2'2'}^+ (h_0(\varepsilon-eV_2)-h_0(\varepsilon))\}
K_{21}^{+^{-1}} \\\times (S_{11'}^+ S_{11'} - |a^2|
S_{1'1'}^{+} \sigma_2 \underline{S}_{1'1} \underline{S}_{1'1}^+
\sigma_2 S_{1'1'}) ], \label{sp_g}
\end{multline}
where we explicitly used the fact that in equilibrium $\Sp
(\hat\tau_3 \hat g^K_{1^{out}} - \hat\tau_3 \hat
g^K_{1^{in}})\equiv 0$. Substituting \eqref{sp_g} into
\eqref{current_g}, we finally obtain
\begin{equation}
I_1= I_1^{BTK}(V_1)+I_{11}(V_1) + I_{12}(V_2).
\label{final}
\end{equation}
The correction to the local BTK current (arising from trajectories
crossing also the second NS interface) has the following form
\begin{multline}
I_{11}(V_1)=
-\dfrac{G_0}{4e} \int d \varepsilon
\bigl\{
\Sp[r_{1l}^R r_{1l}^{R^+} - \tilde \Gamma_1^R  \tilde
\Gamma_1^{R^+} -1] (h_0(\varepsilon+eV_1)-h_0(\varepsilon)) \\-
\Sp[\tilde r_{1l}^R \tilde r_{1l}^{R^+}- \Gamma_1^R \Gamma_1^{R^+}
-1] (h_0(\varepsilon-eV_1)-h_0(\varepsilon)) \bigr\}, \label{I11}
\end{multline}
while for the cross-current we obtain
\begin{multline}
I_{12}(V_2)=
-\dfrac{G_0}{4e} \int d \varepsilon
(1-\tanh^2iL\Omega/v_F)
\\\times
\Sp[K_{21}^{-1}
\{S_{2'2}S_{2'2}^+ (h_0(\varepsilon+eV_2)-h_0(\varepsilon))
\\+
|a|^2 S_{2'2'} \sigma_2 \underline{S}_{22'}^+\underline{S}_{22'} \sigma_2 S_{2'2'}^+
(h_0(\varepsilon-eV_2)-h_0(\varepsilon))\}
K_{21}^{+^{-1}}
\\\times
(S_{11'}^+ S_{11'} -
|a^2| S_{1'1'}^{+} \sigma_2 \underline{S}_{1'1} \underline{S}_{1'1}^+ \sigma_2 S_{1'1'})
].
\label{I12}
\end{multline}
\begin{figure}
\centerline{
\includegraphics[width=85mm]{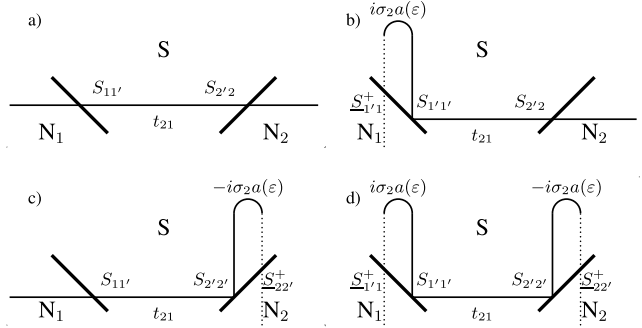}
}
\caption{Diagrams representing four different contributions to the
cross-current $I_{12}$ \eqref{I12}. Solid (dotted) lines correspond to
propagating electron-like (hole-like) excitations and $t_{21}=K_{21}^{-1} / \cosh(iL\Omega/v_F)$.}
\label{traject}
\end{figure}
Eqs. \eqref{final}-\eqref{I12} fully determine the current across
the first interface at arbitrary voltages, temperature and
spin-dependent interface transmissions.

In right hand side of Eq. \eqref{I12} we can distinguish four
contributions with different products of $S$-matrices. Each of
these terms corresponds to a certain sequence of elementary
events, such as transmission, reflection, Andreev reflection and
propagation between interfaces. Diagrammatic representation of
these four terms is offered in Fig.~\ref{traject}. The amplitude
of each of the processes is given by the product of the amplitudes
of the corresponding elementary events. For instance, the
amplitude of the process in Fig.~\ref{traject}c  is $f=-i S_{11'}t_{21}
S_{2'2'} a \sigma_2 \underline{S}_{22'}^+$. In  Eq.\eqref{I12}
this process is identified by the term  $\Sp(f f^+)$ with the hole
distribution function as a prefactor. It is straightforward to
observe that the processes of Fig.~\ref{traject}a, \ref{traject}b 
and \ref{traject}d correspond to the
other three terms in \eqref{I12}. We also note that the processes
of Fig.~\ref{traject}a and \ref{traject}d describe direct electron 
(hole) transport, while
the processes of Fig.~\ref{traject}b and \ref{traject}c correspond 
to the contribution of CAR.

Assuming that both interfaces possess inversion symmetry as well
as reflection symmetry in the plane normal to the corresponding
interface we can choose $\mathcal{S}$-matrices in the following
form
\begin{gather}
S_{11}=S_{1'1'}=\underline{S}^T_{11}=\underline{S}^T_{1'1'}
=
U(\varphi)
\begin{pmatrix}
\sqrt{R_{1\uparrow}}e^{i\theta_1/2} & 0 \\
0 & \sqrt{R_{1\downarrow}}e^{-i\theta_1/2} \\
\end{pmatrix}
U^+(\varphi),
\label{S11}
\\
S_{11'}=S_{1'1}=\underline{S}^T_{11'}=\underline{S}^T_{1'1} =
U(\varphi) i
\begin{pmatrix}
\sqrt{D_{1\uparrow}}e^{i\theta_1/2} & 0 \\
0 & \sqrt{D_{1\downarrow}}e^{-i\theta_1/2} \\
\end{pmatrix}
U^+(\varphi),
\\
S_{22}=S_{2'2'}=\underline{S}_{22}=\underline{S}_{2'2'}
=
\begin{pmatrix}
\sqrt{R_{2\uparrow}}e^{i\theta_2/2} & 0 \\
0 & \sqrt{R_{2\downarrow}}e^{-i\theta_2/2} \\
\end{pmatrix},
\label{S22}
\\
S_{22'}=S_{2'2}=\underline{S}_{22'}=\underline{S}_{2'2}
=
i
\begin{pmatrix}
\sqrt{D_{2\uparrow}}e^{i\theta_2/2} & 0 \\
0 & \sqrt{D_{2\downarrow}}e^{-i\theta_2/2} \\
\end{pmatrix}.
\label{S22'}
\end{gather}
Here $R_{1(2)\uparrow (\downarrow )}=1-D_{1(2)\uparrow (\downarrow
)}$ are the spin dependent reflection coefficients of both NS
interfaces, $\theta_{1,2}$ are spin-mixing angles and $U(\varphi)$
is the rotation matrix in the spin
space which depends on the angle $\varphi$ between polarizations
of the two interfaces,
\begin{equation}
U(\varphi)=\exp(-i\varphi\sigma_1/2)=
\begin{pmatrix}
\cos( \varphi/2) & -i \sin(\varphi/2) \\
-i \sin(\varphi/2) & \cos( \varphi/2) \\
\end{pmatrix}.
\end{equation}

In general spin current is not conserved in heterostructures with spin active interfaces. However single barrier with $S$-matrix \eqref{S22}-\eqref{S22'} does not violate spin current conservation \cite{Zhao07}. It is easy to show that in our two barrier structure with  interface $S$-matrices \eqref{S11}-\eqref{S22'} spin current is conserved in the normal state for arbitrary barriers polarizations and in superconducting state for collinear barriers polarizations.

Substituting the above expressions for the $S$-matrices into Eqs.
\eqref{I11} and \eqref{I12} we arrive at the final results for
both $I_{11}(V_1)$ and $I_{12}(V_2)$ which will be specified
further below.

\subsection{Cross-current}
First let us consider the cross-current $I_{12}(V_2)$. From the above
analysis we obtain
\begin{multline}
I_{12}(V_2)= -\dfrac{G_0}{4e} \int d \varepsilon
\left[\tanh\dfrac{\varepsilon+eV_2}{2T} -
\tanh\dfrac{\varepsilon}{2T} \right]
\dfrac{1-\tanh^2iL\Omega/v_F}{W(z_1,z_2,\varepsilon,\varphi)}
\\  \times \Bigl\{ \left[D_{1\downarrow}D_{2\downarrow}-
|a|^2D_{1\uparrow}D_{2\downarrow}(R_{1\downarrow}+R_{2\uparrow})+
|a|^4
D_{1\downarrow}R_{1\uparrow}D_{2\downarrow}R_{2\uparrow}\right]
|K(z_1,z_2,\varepsilon)|^2\tilde c
\\ +
\left[D_{1\uparrow} D_{2\uparrow}
-|a|^2D_{1\downarrow}D_{2\uparrow}(R_{1\uparrow}+R_{2\downarrow})
+|a|^4 D_{1\uparrow}
R_{1\downarrow}D_{2\uparrow}R_{2\downarrow}\right]
|K(z_1^*,z_2^*,\varepsilon)|^2\tilde c
\\ +
\left[D_{1\uparrow}D_{2\downarrow}-
|a|^2D_{1\downarrow}D_{2\downarrow}(R_{1\uparrow}+R_{2\uparrow})+
|a|^4
D_{1\uparrow}R_{1\downarrow}D_{2\downarrow}R_{2\uparrow}\right]
|K(z_1^*,z_2,\varepsilon)|^2\tilde s
\\ +
\left[D_{1\downarrow} D_{2\uparrow}
-|a|^2D_{1\uparrow}D_{2\uparrow}(R_{1\downarrow}+R_{2\downarrow})
+|a|^4 D_{1\downarrow}
R_{1\uparrow}D_{2\uparrow}R_{2\downarrow}\right]
|K(z_1,z_2^*,\varepsilon)|^2\tilde s \Bigr\},
\label{I12phi}
\end{multline}
where we define $\tilde c=\cos^2(\varphi/2)$, $\tilde s=\sin^2(\varphi/2)$,
\begin{gather}
K(z_1,z_2,\varepsilon)= (1-a^2 z_1 z_2) - \left[ \varepsilon (1+a^2 z_1 z_2) + \Delta a (z_1+z_2) \right]Q,
\\
\begin{split}
W(z_1,z_2,\varepsilon,\varphi)= |K(z_1,z_2,\varepsilon)
& K(z_1^*,z_2^*,\varepsilon) \cos^2(\varphi/2)
\\+
& K(z_1^*,z_2,\varepsilon) K(z_1,z_2^*,\varepsilon)
\sin^2(\varphi/2)|^2
\end{split}
\end{gather}
and $z_i=\sqrt{ R_{i\uparrow} R_{i\downarrow} }\exp( i \theta_i)$
( $i=1,2$).

Eq.~\eqref{I12phi} represents our central result. It fully
determines the non-local spin-dependent current in our
three-terminal ballistic NSN structure at arbitrary voltages,
temperature, interface transmissions and polarizations.

Let us introduce the non-local differential conductance
\begin{equation}
G_{12}(V_2)=-\dfrac{\partial I_{1}}{\partial V_2}
=-\dfrac{\partial I_{12}(V_2)}{\partial V_2}.
\end{equation}
Before specifying this quantity further it is important to observe
that in general the conductance $G_{12}(V_2)$ is not an even
function of the applied voltage $V_2$. This asymmetry arises due
to formation of Andreev bound states in the vicinity of a
spin-active interface \cite{Fogel00,Barash02}. It disappears
provided the spin mixing angles $\theta_1$ and $\theta_2$ remain
equal to $0$ or $\pi$.

In the normal state we have $I_{12}(V_2)=-G_{N_{12}}V_2$, where
\begin{multline}
G_{N_{12}}= \dfrac{G_0}{2} \bigl[ (D_{1\downarrow}D_{2\downarrow}
+ D_{1\uparrow} D_{2\uparrow})\cos^2(\varphi/2) \\+
(D_{1\uparrow}D_{2\downarrow} + D_{1\downarrow}
D_{2\uparrow})\sin^2(\varphi/2) \bigr]. \label{NNN}
\end{multline}

Turning to the superconducting state, let us consider the limit of
low temperatures and voltage $T, V_2 \ll \Delta$. In this
limit only subgap quasiparticles contribute to the cross-current and the
differential conductance becomes voltage-independent, i.e.
$I_{12}=-G_{12}V_{2}$, where
\begin{multline}
G_{12}=G_0(1-\tanh^2L\Delta/v_F)
\\\times
\Biggl\{
\dfrac{D_{1\uparrow}D_{1\downarrow}D_{2\uparrow}D_{2\downarrow}}{
|K(z_1,z_2,0)|^2\tilde c+|K(z_1,z_2^*,0)|^2\tilde s}
\\+
(D_{1\uparrow}-D_{1\downarrow})(D_{2\uparrow}-D_{2\downarrow})
\dfrac{|K(z_1,z_2,0)|^2\tilde c-|K(z_1,z_2^*,0)|^2\tilde s
}{\left(|K(z_1,z_2,0)|^2\tilde c+|K(z_1,z_2^*,0)|^2\tilde s\right)^2}
\Biggr\}.
\label{G12zeroV}
\end{multline}
where, as before, $\tilde c=\cos^2(\varphi/2)$ and $\tilde s=\sin^2(\varphi/2)$.
In the case of spin-isotropic interfaces Eqs. \eqref{G12zeroV} and
\eqref{I12phi} reduce to the results \cite{KZ06}.

\begin{figure}
\centerline{
\includegraphics[width=75mm]{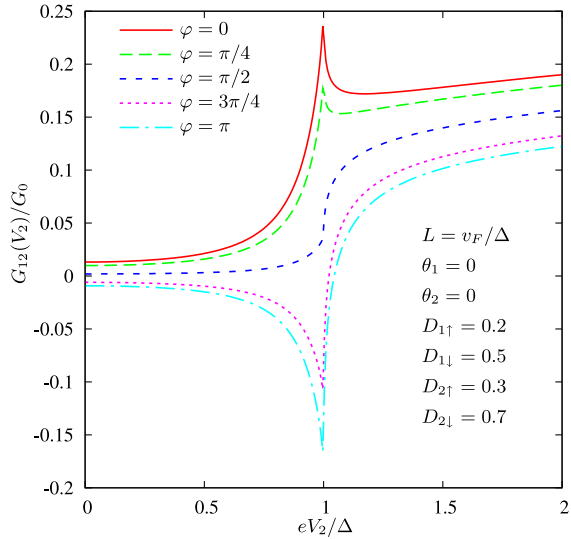}
} \caption{Zero temperature differential non-local conductance as
a function of voltage at zero spin-mixing angles $\theta_{1,2}=0$. }
\label{current-gen-theta=0}
\end{figure}

Provided at least one of the interfaces is spin-isotropic, the
conductance \eqref{G12zeroV} is proportional to the product of all
four transmissions
$D_{1\uparrow}D_{1\downarrow}D_{2\uparrow}D_{2\downarrow}$, i.e.
it differs from zero only due to processes involving scattering
with both spin projections at both NS interfaces. As in the case
of spin-isotropic interfaces \cite{KZ06} the value $G_{12}$
\eqref{G12zeroV} gets strongly suppressed with increasing $L$, and
at sufficiently high interface transmissions this dependence is in
general non-exponential in $L$. In the spin-degenerate case for a
given $L$ the non-local conductance reaches its maximum for
reflectionless barriers $D_{1,2}=1$. In this case we arrive at a
simple formula
\begin{equation}
G_{12}=G_0(1-\tanh^2L\Delta/v_F).
\label{D=1}
\end{equation}
We observe that for small $L \ll v_F/\Delta$ the conductance
$G_{12}$ identically coincides with its normal state value
$G_{N_{12}}\equiv G_0$ at any temperature and voltage \cite{KZ06}.
This result implies that CAR {\it vanishes for fully open
barriers}. Actually this conclusion is general and applies not
only for small but for any value of $L$, i.e. the result
(\ref{D=1}) is determined solely by the process of direct electron
transfer between N-terminals for all $L$.

At the first sight, this result might appear counterintuitive
since the behavior of ordinary (local) AR is just the opposite: It
reaches its maximum at full barrier transmissions. The physics
behind vanishing of CAR for perfectly trasparent NS interfaces is
simple. One observes (cf. Fig.~\ref{car-fig}) that CAR inevitably implies the
flow of Cooper pairs out of the contact area into the
superconducting terminal. This flow is described by electron
trajectories which end deep in the superconductor. On the other
hand, it is obvious that CAR requires ``mixing'' of these
trajectories with those going straight between two normal
terminals. Provided there exists no normal electron reflection at
both NS interfaces such mixing does not occur, CAR vanishes and
the only remaining contribution to the non-local conductance is
one from direct electron transfer between N-terminals.

This situation is illustrated by the diagrams in Fig.~\ref{traject}.
It is obvious that in the case of non-reflecting NS interfaces
only the process of Fig.~\ref{traject}a survives, whereas all other processes
(Fig.~\ref{traject}b, \ref{traject}c and \ref{traject}d) vanish for reflectionless barriers with
$R_{1(2)\uparrow(\downarrow)}=0$. The situation changes provided
at least one of the transmissions is smaller than one. In this
case scattering at SN interfaces mixes up trajectories connecting
N$_1$ and N$_2$ terminals with ones going deep into and coming
from the superconductor. As a result, all four processes depicted
in Fig.~\ref{traject} contribute to the cross-current and CAR contribution to
$G_{12}$ does not vanish.

In the limit $|eV_2|, T \ll \Delta$ and at zero spin-mixing angles
$\theta_{1,2}=0$ from Eq. \eqref{G12zeroV} we obtain
\begin{multline}
G_{12}=G_0\dfrac{1-\tanh^2L\Delta/v_F}{|K(z_1,z_2,0)|^2}
\bigl\{
D_{1\uparrow}D_{1\downarrow}D_{2\uparrow}D_{2\downarrow}
\\+
(D_{1\uparrow}-D_{1\downarrow})(D_{2\uparrow}-D_{2\downarrow})\cos\varphi
\bigr\}.
\label{G12zeroVzeroTheta}
\end{multline}
In the lowest (first order) order in the transmissions of both
interfaces and for collinear interface polarizations Eq.
\eqref{G12zeroVzeroTheta} reduces to the result by Falci {\it et
al.} \cite{FFH} provided we identify the tunneling density of
states $N_0 D_{1\uparrow}$, $N_0 D_{1\downarrow}$, $N_0
D_{2\uparrow}$, and $N_0 D_{2\downarrow}$ with the corresponding
spin-resolved densities of states in the ferromagnetic electrodes.
For zero spin-mixing angles and low voltages the $L$-dependence of
the nonlocal conductance $G_{12}$ reduces to the exponential form
$G_{12} \propto \exp (-2L\Delta /v_F)$ either in the limit of
small transmissions or large  $L \gg v_F/\Delta$.

\begin{figure}
\centerline{
\includegraphics[width=75mm]{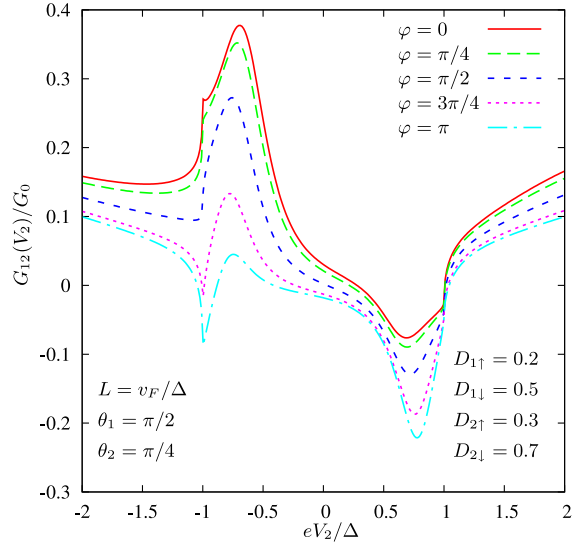}
} \caption{The same as in Fig.~\ref{current-gen-theta=0} for $\theta_1=\pi/2$, $\theta_2=\pi/4$. }
\label{current-gen-theta-1}
\end{figure}

At arbitrary voltages and temperatures the cross-current has a
simple $\varphi$ dependence in the limit of zero spin mixing
angles ($\theta_{1,2}=0$)
\begin{equation}
I_{12}(\varphi,V_2)= I_{12}(\varphi=0,V_2)\cos^2(\varphi/2) +
I_{12}(\varphi=\pi,V_2)\sin^2(\varphi/2), \label{I12theta0}
\end{equation}
i.e. in this limit at any $\varphi$ the nonlocal current is equal
to a proper superposition of the two contributions corresponding
to parallel ($\varphi=0$) and antiparallel ($\varphi=\pi$)
interface polarizations. Some typical curves for the differential
non-local conductance are presented in
Fig.~\ref{current-gen-theta=0} at sufficiently high interface
transmissions and zero spin mixing angles $\theta_{1,2}=0$.

Let us now turn to the limit of highly polarized interfaces which
is accounted for by taking the limit of vanishing spin-up (or
spin-down) transmission of each interface. In this limit our model
describes an HSH structure, where H stands for fully
spin-polarized half-metallic electrodes. In this case we obtain
($D_{1\uparrow}=D_1$, $D_{1\downarrow}=0$, $D_{2\uparrow}=D_2$,
and $D_{2\downarrow}=0$)
\begin{multline}
I_{12}(V_2)=
-\dfrac{G_0}{4e} \int d \varepsilon
\left[h_0(\varepsilon+eV_2) -h_0(\varepsilon)\right]
\dfrac{1-\tanh^2iL\Omega/v_F}{W(z_1,z_2,\varepsilon,\varphi)}
D_1 D_2
\\\times
\Bigl\{
\left[1+|a|^4\right]|K(z_1^*,z_2^*,\varepsilon)|^2\tilde c
-
2 |a|^2|K(z_1,z_2^*,\varepsilon)|^2\tilde s
\Bigr\}.
\label{hsh}
\end{multline}
We observe that the nonlocal conductance has {\it opposite signs} for
parallel ($\varphi=0$) and antiparallel ($\varphi=\pi$) interface
polarizations. We also emphasize that, as it is also clear from
Eq. (\ref{G12zeroVzeroTheta}), the cross-conductance $G_{12}$
of HSH structures -- in contrast to that for NSN structures --
does not vanish already in the lowest order in barrier transmissions
$D_{1\uparrow}D_{2\uparrow}$.

In general the non-local conductance is very sensitive to
particular values of the spin-mixing angles $\theta_1$ and
$\theta_2$, as illustrated, e.g., in Fig.~\ref{current-gen-theta-1}. 
Comparing the voltage dependencies of
the nonlocal conductance evaluated for the same transmissions and
presented in Figs.\ref{current-gen-theta=0} and
\ref{current-gen-theta-1}, we observe that they can differ
drastically at zero and non-zero values of $\theta_{1,2}$.

At low voltages and temperatures and at zero spin mixing angles
the non-local conductance of HSH structures is determined by Eq.
(\ref{G12zeroVzeroTheta}) with
$D_{1\downarrow}=D_{2\downarrow}=0$. For fully open barriers (for
''spin-up'' electrons) $D_{1\uparrow}=D_{2\uparrow}=1$ we obtain
\begin{equation}
G_{12}=G_0(1-\tanh^2L\Delta/v_F)\cos \varphi . \label{DD=1}
\end{equation}
Interestingly, for $\varphi =0$ this expression exactly coincides
with that for fully open NSN structures, Eq. (\ref{D=1}). At the
same time for small $L$ the result (\ref{DD=1}) turns out to be 2
times bigger that the analogous expression in the normal case, i.e. for
fully open HNH structures, cf. Eq. (\ref{NNN}). This result can
easily be interpreted in terms of diagrams in Fig.~\ref{traject}. We observe that
-- exactly as for the spin degenerate case -- CAR diagrams of Fig.~\ref{traject}b,c vanish for reflectionless barriers, whereas diagrams of
Fig.~\ref{traject}a,d describing direct electron transfer survive and both
contribute to $G_{12}$. Thus, {\it CAR vanishes identically also
for fully open HSH structures}. The factor of 2 difference with
the normal case is due to the fact that the diagram of Fig.~\ref{traject}d
vanishes in the normal limit.

\subsection{Correction to BTK}
Using the above formalism one can easily generalize the BTK result
to the case of spin-polarized interfaces\cite{Zhao04}.
For the first interface we have
\begin{multline}
I_1^{BTK}(V_1)= \dfrac{\mathcal{N}_1}{R_qe} \int d \varepsilon
[h_0(\varepsilon+eV_1)-h_0(\varepsilon)] (1+|a|^2) \\\times
\left< \dfrac{|v_{x_1}|}{v_F} \left(
D_{1\uparrow}\dfrac{1-R_{1\downarrow}|a|^2}{|1-z_1 a^2|^2} +
D_{1\downarrow}\dfrac{1-R_{1\uparrow}|a|^2}{|1-z_1^* a^2|^2}
\right) \right>.
\end{multline}
Here transmission and reflection coefficients as well as the spin
mixing angle depend on the direction of the Fermi momentum. In the
spin-degenerate case the above expression reduces to the standard
BTK result \cite{BTK}.

Evaluating the nonlocal correction to the BTK current due to the
presence of the second interface we arrive at a somewhat lengthy
general expression
\begin{multline}
I_{11}(V_1)=
\dfrac{G_0}{2e} \int d \varepsilon
(h_0(\varepsilon+eV_1)-h_0(\varepsilon))
\dfrac{1}{W(z_1,z_2,\varepsilon,\varphi)}\Bigl\{ 2W(z_1,z_2,\varepsilon,\varphi)
\\-
R_{1\uparrow}
\bigl| \tilde c K(z_1/R_{1\uparrow}, z_2,\varepsilon) K(z_1^*, z_2^*,\varepsilon) +
\tilde s K(z_1/R_{1\uparrow}, z_2^*,\varepsilon) K(z_1^*, z_2,\varepsilon) \bigr|^2
\\-
R_{1\downarrow}
\bigl| \tilde c K(z_1^*/R_{1\downarrow}, z_2^*,\varepsilon) K(z_1, z_2,\varepsilon) +
\tilde s K(z_1^*/R_{1\downarrow}, z_2,\varepsilon) K(z_1, z_2^*,\varepsilon) \bigr|^2
\Bigr\}
\\
+\dfrac{G_0}{4e} \int d \varepsilon
(h_0(\varepsilon+eV_1)-h_0(\varepsilon))
\dfrac{D_{1\uparrow}D_{1\downarrow}}{W(z_1,z_2,\varepsilon,\varphi)}\Bigl\{
\\
|a|^2 \bigl| \tilde c K(0,z_2,\varepsilon) K(z_1^*, z_2^*,\varepsilon) +
\tilde s K(0,z_2^*,\varepsilon) K(z_1^*, z_2,\varepsilon) \bigr|^2
\\+
|a|^2 \bigl| \tilde c K(0,z_2^*,\varepsilon) K(z_1, z_2,\varepsilon) +
\tilde s K(0,z_2,\varepsilon) K(z_1, z_2^*,\varepsilon) \bigr|^2
\\+
\dfrac{1}{|a|^2}
\bigl| \tilde c K'(z_2^*,\varepsilon) K(z_1, z_2,\varepsilon) +
\tilde s K'(z_2,\varepsilon) K(z_1, z_2^*,\varepsilon) \bigr|^2
\\+
\dfrac{1}{|a|^2}
\bigl| \tilde c K'(z_2,\varepsilon) K(z_1^*, z_2^*,\varepsilon) +
\tilde s K'(z_2^*,\varepsilon) K(z_1^*, z_2,\varepsilon) \bigr|^2
\Bigr\}
\\+
\dfrac{G_0}{e} R_{2\uparrow}R_{2\downarrow}\sin^2(\theta_2/2)
\tilde s\tilde c
\\\times
\int d \varepsilon
(h_0(\varepsilon+eV_1)-h_0(\varepsilon))
\dfrac{(1-\tanh^2iL\Omega/v_F)^2
}{W(z_1,z_2,\varepsilon,\varphi)}
\\\times
\left[|a|^2(D^2_{1\uparrow}+D^2_{1\downarrow}) - 2 |a|^4
D_{1\uparrow}D_{1\downarrow}(R_{1\uparrow} + R_{1\downarrow}) +
|a|^6
(D^2_{1\uparrow}R^2_{1\downarrow}+D^2_{1\downarrow}R^2_{1\uparrow})\right],
\label{CARcorr}
\end{multline}
where $K'(z_2,\varepsilon)=\partial K(z_1,
z_2,\varepsilon)/\partial z_1$. This expression gets significantly
simplified in the limit of zero spin-mixing angles $\theta_{1,2}=
0$ in which case we obtain
\begin{multline}
I_{11}(V_1)=
\dfrac{G_0}{2e} \int d \varepsilon
(h_0(\varepsilon+eV_1)-h_0(\varepsilon))
\\\times
\Biggl\{
2-
R_{1\uparrow}
\dfrac{|K(z_1/R_{1\uparrow}, z_2,\varepsilon)|^2}{|K(z_1,z_2,\varepsilon)|^2}
-
R_{1\downarrow}
\dfrac{|K(z_1/R_{1\downarrow}, z_2,\varepsilon)|^2}{|K(z_1,z_2,\varepsilon)|^2}
\\+
D_{1\uparrow}D_{1\downarrow}
\dfrac{
|a(\varepsilon)|^2 |K(0,z_2,\varepsilon)|^2+
|K'(z_2,\varepsilon)|^2/|a(\varepsilon)|^2}{|K(z_1,z_2,\varepsilon)|^2}
\Biggr\}.
\end{multline}
In contrast to the expression for the cross-current $I_{12}$ (cf.
Eq. \eqref{I12theta0}), in the limit of zero spin-mixing angles
the correction $I_{11}$ to the BTK current does not depend on the
angle $\varphi$ between the interface polarizations. In
particular, at $|eV_1|, T \ll \Delta$ we have $I_{11}= G_{11}V_1$
where
\begin{multline}
G_{11}=
G_0
(D_{1\uparrow}+D_{1\downarrow})
\dfrac{(1-z_2^2)(1-\tanh^2L\Delta/v_F)}{
[1+z_1 z_2 +(z_1+z_2 )\tanh L\Delta/v_F]^2}
\\+
G_0
D_{1\uparrow}D_{1\downarrow}
\dfrac{(1+z_2 \tanh L\Delta/v_F)^2 + 3 (z_2 + \tanh L\Delta/v_F)^2}{[1+z_1 z_2 +(z_1+z_2 )\tanh L\Delta/v_F]^2}.
\end{multline}
In the tunneling limit
$D_{1\uparrow},D_{1\downarrow},D_{2\uparrow},D_{2\downarrow} \ll
1$ we reproduce the result of Ref. \cite{FFH}
\begin{equation}
G_{11}=\dfrac{G_0}{4}
(D_{1\uparrow}+D_{1\downarrow})(D_{2\uparrow}+D_{2\downarrow})
\exp(-2L\Delta/v_F),
\end{equation}
which turns out to hold at any value $\varphi$.

As compared to the BTK conductance the CAR correction
\eqref{CARcorr} contains an extra small factor ${\cal A}_2/L^2$
and, hence, in many cases remains small and can be neglected. On
the other hand, since CAR involves tunneling of {\it one} electron
through each interface, for strongly asymmetric structures with
$D_{1\uparrow},D_{1\downarrow} \ll 1$ and $D_{2\uparrow},
D_{2\downarrow} \sim 1$ it can actually {\it strongly exceed} the
BTK conductance. Indeed, for $D_{1\uparrow\downarrow} \ll 1$,
$R_{2\uparrow}R_{2\downarrow} \ll 1$ and provided the spin mixing
angle $\theta_1$ is not very close to $\pi$ from Eq.
\eqref{CARcorr} we get
\begin{equation}
G_{11}= \dfrac{G_0 (D_{1\uparrow}+D_{1\downarrow})}{\cosh (2
L\Delta/v_F) + \cos \theta_1 \sinh (2 L\Delta/v_F)},
\label{CARcorr1}
\end{equation}
i.e. for
$$
\frac{D_{1\uparrow} D_{1\downarrow}}{(D_{1\uparrow} +
D_{1\downarrow})} < \frac{{\cal A}_2}{L^2}\exp(-2L\Delta/v_F)
$$
the contribution \eqref{CARcorr1} may well exceed the BTK term
$G_{1}^{BTK} \propto D_{1\uparrow} D_{1\downarrow}$. The existence
of such a non-trivial regime further emphasizes the importance of
the mechanism of non-local Andreev reflection in multi-terminal
hybrid NSN structures.

\section{Diffusive FSF structures}

\begin{figure}
\centerline{\includegraphics[width=7.5cm]{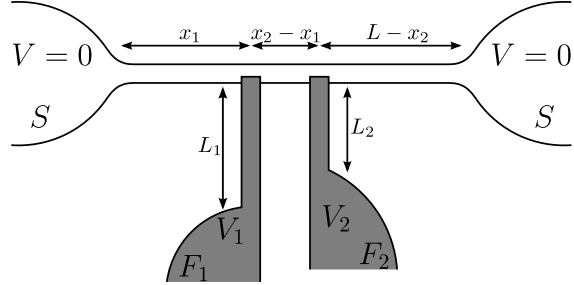}}
\caption{FSF structure under consideration.}
\label{ufsf-fig}
\end{figure}

Let us now turn to the effect of disorder. In what follows we
will consider a three-terminal diffusive FSF structure
schematically shown in Fig.~\ref{ufsf-fig}. Two ferromagnetic terminals F$_1$
and F$_2$ with resistances $r_{N_1}$ and $r_{N_2}$ and electric
potentials $V_1$ and $V_2$ are connected to a superconducting
electrode of length $L$ with normal state (Drude) resistance $r_L$
and electric potential $V=0$ via tunnel barriers. The magnitude of
the exchange field $h_{1,2}=|\bm{h}_{1,2}|$ in both ferromagnets F$_1$ and
F$_2$ is assumed to be much bigger than the superconducting order
parameter $\Delta$ of the S-terminal and, on the other hand,
much smaller that the Fermi energy, i.e. $\Delta \ll h_{1,2} \ll
\epsilon_F$. The latter condition allows to perform the analysis of our FSF
system within the quasiclassical formalism of Usadel equations for
the Green-Keldysh matrix functions $G$ formulated below.

\subsection{Quasiclassical equations}

In each of our metallic
terminals the Usadel equations can be written in the form \cite{BWBSZ}
\begin{equation}
iD\nabla (\check G \nabla \check G)=
[\check\Omega +eV , \check G], \quad \check G^2=1,
\label{Usadel}
\end{equation}
where $D$ is the diffusion constant, $V$ is the electric potential, $\check G$ and $\check\Omega$ are $8\times8$ matrices in Keldysh-Nambu-spin space (denoted by check symbol)
\begin{gather}
\check G=
\begin{pmatrix}
\breve G^R & \breve G^K \\
0 & \breve G^A \\
\end{pmatrix}, \quad
\check \Omega=
\begin{pmatrix}
\breve \Omega^R & 0 \\
0 & \breve \Omega^A \\
\end{pmatrix},
\\
\breve \Omega^R=\breve \Omega^A=
\begin{pmatrix}
\varepsilon - \hat{\bm{\sigma}}\bm{h} & \Delta \\
-\Delta^* & -\varepsilon + \hat{\bm{\sigma}}\bm{h}\\
\end{pmatrix},
\end{gather}
$\varepsilon$ is the quasiparticle energy, $\Delta (T)$ is
the superconducting order parameter which will be considered real
in a superconductor and zero in both ferromagnets, $\bm{h}\equiv\bm{h}_{1(2)}$ in the
first (second) ferromagnetic terminal, $\bm{h}\equiv 0$ outside these terminals and $\hat{\bm{\sigma}}=(\hat\sigma_1,\hat\sigma_2,\hat\sigma_3)$ are Pauli matrices in spin space.

Retarded and advanced Green functions $\breve G^R$ and $\breve G^A$ have the following matrix structure
\begin{equation}
\breve G^{R,A}=
\begin{pmatrix}
\hat G^{R,A}  & \hat F^{R,A} \\
-\hat F^{R,A} & -\hat G^{R,A} \\
\end{pmatrix}.
\end{equation}
Here and below $2\times 2$ matrices in spin space are denoted by hat symbol.

Having obtained the expressions for the Green-Keldysh functions $\check G$ one can easily evaluate the current density $\bm{j}$ in our system
with the aid of the standard relation
\begin{equation}
\bm{j}= -\frac{\sigma}{16e} \int \Sp [ \tau_3 ( \check G \nabla  \check G)^K]d
\varepsilon ,\label{current_dif}
\end{equation}
where $\sigma$ is the Drude conductivity of the corresponding
metal and $\tau_3$ is the Pauli matrix in Nambu space.

In what follows it will be convenient for us to employ the
so-called Larkin-Ovchinnikov parameterization of the Keldysh Green
function
\begin{equation}
\breve G^K=\breve G^R \breve f - \breve f \breve G^A,
\quad \breve f = \hat f_L + \tau_3 \hat f_T,
\end{equation}
where the distribution functions $\hat f_L$ and $\hat f_T$ are $2\times2$
matrices in the spin space.

For the sake of simplicity we will assume that
magnetizations of both ferromagnets and the interfaces (see below)
are collinear. Within this approximation the Green functions and
the matrix $\check \Omega$ are diagonal in the spin space and the
diffusion-like equations for the distribution function matrices
$\hat f_L$ and $\hat f_T$ take the form
\begin{gather}
- D \nabla \left( \hat D^T(\bm{r},\varepsilon) \nabla
\hat f_T(\bm{r},\varepsilon) \right) + 2\hat \Sigma(\bm{r},\varepsilon)
\hat f_T(\bm{r},\varepsilon) =0, \label{diffeqT}
\\
- D \nabla \left( \hat D^L(\bm{r},\varepsilon) \nabla
\hat f_L(\bm{r},\varepsilon) \right) =0, \label{diffeqL}
\end{gather}
where
\begin{gather}
\hat \Sigma (\bm{r},\varepsilon)= - i \Delta \Img \hat F^R,
\\
\hat D^T =\left( \Rea \hat G^R\right)^2 + \left( \Img \hat F^R\right)^2,
\\
\hat D^L =\left( \Rea \hat G^R\right)^2 - \left( \Rea \hat F^R\right)^2.
\end{gather}
The function $\hat \Sigma (\bm{r},\varepsilon)$ differs from zero only
inside the superconductor. It accounts both for energy relaxation of
quasiparticles and for their conversion to Cooper pairs due to Andreev
reflection. The functions $\hat D^T$ and $\hat D^L$ acquire space and energy
dependencies due to the presence of the superconducting wire and
renormalize the diffusion coefficient $D$.

The solution of Eqs. \eqref{diffeqT}-\eqref{diffeqL} can be
expressed in terms of the diffuson-like functions $\hat{\mathcal{D}}^T$
and $\hat{\mathcal{D}}^L$ which obey the following equations
\begin{gather}
- D \nabla \left[
\hat D^T(\bm{r},\varepsilon) \nabla \hat{\mathcal{D}}^T(\bm{r},\bm{r}^{\prime},\varepsilon)
\right]
+2\hat \Sigma (\bm{r},\varepsilon)
\hat{\mathcal{D}}^T(\bm{r},\bm{r}^{\prime},\varepsilon) =
\delta(\bm{r}-\bm{r}^{\prime}),
\\
- D \nabla \left[
\hat D^L(\bm{r},\varepsilon) \nabla \hat{\mathcal{D}}^L(\bm{r},\bm{r}^{\prime},\varepsilon)
\right] =
\delta(\bm{r}-\bm{r}^{\prime}).
\end{gather}

\subsection{Boundary conditions}

The solutions of Usadel equation (\ref{Usadel}) in each of the metals
should be matched at FS-interfaces by means of appropriate
boundary conditions which account for electron tunneling between
these terminals. The form of these boundary conditions essentially
depends on the adopted model describing electron scattering at
FS-interfaces. As before, we stick to the model of the so-called
spin-active interfaces which takes into account
possibly different barrier transmissions for spin-up and spin-down
electrons. Here we
employ this model in the case of diffusive electrodes and also
restrict our analysis to the case of tunnel barriers with channel
transmissions much smaller than one. In this case the
corresponding boundary conditions read \cite{Huertas02,Huertas09}
\begin{gather}
\mathcal{A}\sigma_+  \check G_+  \partial_x \check G_+=
\dfrac{G_T}{2}
[\check G_-, \check G_+]
+\dfrac{G_m}{4}
[\{\hat{\bm{\sigma}}\bm{m} \tau_3, \check G_-\}, \check G_+] +
i\dfrac{G_{\varphi}}{2}
[\hat{\bm{\sigma}}\bm{m} \tau_3, \check G_+],
\label{gplus}
\\
-\mathcal{A}  \sigma_-  \check G_-  \partial_x \check G_-=
\dfrac{G_T}{2}
[\check G_+, \check G_-]
+\dfrac{G_m}{4}
[\{\hat{\bm{\sigma}}\bm{m} \tau_3, \check G_+\}, \check G_-] +
i\dfrac{G_{\varphi}}{2}
[\hat{\bm{\sigma}}\bm{m} \tau_3, \check G_-],
\label{gminus}
\end{gather}
where $\check G_-$ and $\check G_+$ are the Green-Keldysh functions from the left ($x<0$) and from the right ($x>0$) side of the interface, $\mathcal{A}$ is the effective contact
area, $\bm{m}$ is the unit vector in the direction of the
interface magnetization, $\sigma_{\pm}$ are Drude
conductivities of the left and right terminals and $G_T$ is the
spin-independent part of the interface conductance. Along with
$G_T$ there also exists the spin-sensitive contribution to the
interface conductance which is accounted for by the $G_m$-term. The value $G_m$ equals to the difference between interface conductances for spin-up and spin-down conduction bands in the normal state. The $G_{\varphi}$-term arises due to different phase
shifts acquired by scattered quasiparticles with opposite spin
directions.

Employing the above boundary conditions we can establish the
following linear relations between the distribution functions at
both sides of the interface
\begin{gather}
\mathcal{A}\sigma_+ \hat D_+^T \partial_x \hat f_{+T} =\mathcal{A}\sigma_-
\hat D_-^T \partial_x \hat f_{-T}
=\hat g_T (\hat f_{+T} - \hat f_{-T}) + \hat g_m (\hat f_{+L} - \hat f_{-L}) ,\label{boundT}
\\
\mathcal{A}\sigma_+ \hat D_+^L \partial_x \hat f_{+L}=\mathcal{A}\sigma_-
\hat D_-^L \partial_x \hat f_{-L}
=\hat g_L (\hat f_{+L} - \hat f_{-L}) + \hat g_m (\hat f_{+T} - \hat f_{-T}) ,\label{boundL}
\end{gather}
where $\hat g_T$, $\hat g_L$, and $\hat g_m$ are matrix interface conductances which
depend on the retarded and advanced Green functions at the
interface
\begin{gather}
\hat g_T= G_T \left[\left(\Rea \hat G^R_+\right) \left(\Rea \hat G^R_-\right)
+ \left(\Img \hat F^R_+\right)\left(\Img \hat F^R_-\right)\right],
\\
\hat g_L= G_T \left[ \left(\Rea \hat G^R_+\right) \left(\Rea \hat G^R_-\right) -
\left(\Rea \hat F^R_+\right)\left(\Rea \hat F^R_-\right)\right],
\\
\hat g_m = G_m \hat{\bm{\sigma}}\bm{m}\left(\Rea \hat G^R_+\right) \left(\Rea
\hat G^R_-\right).
\end{gather}
Note that the above boundary conditions for the distribution functions do not contain the $G_{\varphi}$-term explicitly since this term in Eqs.
\eqref{gplus}-\eqref{gminus} does not mix Green functions from both sides of the interface.

The current density \eqref{current_dif} can then be expressed in terms
of the distribution function $\hat f_T$ as
\begin{equation}
\bm{j}= -\frac{\sigma}{4e} \int \Sp [ \hat D^T  \nabla  \hat f_T ]d
\varepsilon . \label{current1}
\end{equation}

\subsection{Spectral conductances}
Let us now employ the above formalism in order to evaluate electric currents in our
FSF device. The current across the first (SF$_1$) interface can be written as
\begin{multline}
I_1 = \dfrac{1}{e}\int g_{11}(\varepsilon)\left[f_0(\varepsilon
+ eV_1)-f_0(\varepsilon)\right] d \varepsilon
\\
-\dfrac{1}{e}\int g_{12}(\varepsilon)\left[f_0(\varepsilon +
eV_2)-f_0(\varepsilon)\right] d \varepsilon, \label{intcur1}
\end{multline}
where $f_0(\varepsilon)=\tanh(\varepsilon/2T)$, $g_{11}$ and
$g_{12}$ are local and nonlocal spectral electric conductances.
Expression for the current across the second interface can be
obtained from the above equation by interchanging the indices
$1\leftrightarrow2$. Solving Eqs. \eqref{diffeqT}-\eqref{diffeqL}
with boundary conditions \eqref{boundT}-\eqref{boundL} we express
both local and nonlocal conductances $\hat g_{ij}(\varepsilon)$ in terms of
the interface conductances and the function $\hat{\mathcal{D}}$. The
corresponding results read
\begin{gather}
\hat g_{11}(\varepsilon)= \bigl(
\hat R_2^T \hat{\mathcal{M}}^{L}+
\hat R_2^T \hat R_2^L \hat R_{1m} - \hat R_1^L \hat R_{2m}^2
+ \hat R_{12}^{T} \hat R_{12}^{L} \hat R_{2m}-
\hat R_{1m} \hat R_{2m}^2
\bigr)\hat{\mathcal{K}},
\label{g11}
\\
\hat g_{12}(\varepsilon) = \hat g_{21}(\varepsilon) =\bigl(
\hat R_{12}^T \hat {\mathcal{M}}^{L}+
\hat R_2^T \hat R_{12}^L \hat R_{1m}
+
\hat R_{12}^{L} \hat R_{1m} \hat R_{2m} +
\hat R_{12}^{T} \hat R_{1}^L \hat R_{2m}
\bigr)\hat{\mathcal{K}},
\label{g12}
\end{gather}
where we defined
\begin{gather}
\hat{\mathcal{M}}^{T,L}=\hat R_1^{T,L} \hat R_2^{T,L}-(\hat R_{12}^{T,L})^2,
\\
\hat{\mathcal{K}}^{-1}=
\hat{\mathcal{M}}^{T} \hat{\mathcal{M}}^{L}+
\hat R_{1m}^2 \hat R_{2m}^2 -
\hat R_2^T \hat R_2^L \hat R_{1m}^2
-
2 \hat R_{12}^{T} \hat R_{12}^{L} \hat R_{1m} \hat R_{2m} -
\hat R_1^T \hat R_1^L \hat R_{2m}^2
\label{calK}
\end{gather}
and introduced the auxiliary resistance matrix
\begin{multline}
\hat R_1^T=
\hat g_{1T}(\varepsilon)[
\hat g_{1T}(\varepsilon)\hat g_{1L}(\varepsilon) - \hat g_{1m}^2(\varepsilon)]^{-1}
\\+
\dfrac{D_1 \hat{\mathcal{D}}_1^T(\bm{r}_1,\bm{r}_1,\varepsilon)}{\sigma_1} +
\dfrac{D_S \hat{\mathcal{D}}_S^T(\bm{r}_1,\bm{r}_1,\varepsilon)}{\sigma_S},
\label{RT}
\end{multline}
The resistance matrices $\hat R_2^T$, $\hat R_1^L$ and $\hat R_2^L$ can be obtained by interchanging the indices $1\leftrightarrow2$ and $T\leftrightarrow L$ in Eq. (\ref{RT}). The remaining resistance matrices $\hat R_{12}^{T,L}$ and $\hat R_{jm}$ are defined as
\begin{gather}
\hat R_{12}^{T,L}= \hat R_{21}^{T,L}=
\dfrac{D_S \hat{\mathcal{D}}_S^{T,L}(\bm{r}_1,\bm{r}_2,\varepsilon)}{\sigma_S},
\\
\hat R_{jm}=\hat g_{jm}(\varepsilon)[
\hat g_{jT}(\varepsilon)\hat g_{jL}(\varepsilon) - \hat g_{jm}^2(\varepsilon)]^{-1},
\label{Rm}
\end{gather}
where $j=1,2$. The spectral conductance $g_{ij}$ can be recovered from
the matrix $\hat g_{ij}$ simply by summing up over the spin states
\begin{equation}
g_{ij}(\varepsilon)=\dfrac{1}{2}\Sp\left[\hat g_{ij}(\varepsilon)\right].
\end{equation}

It is worth pointing out that Eqs. \eqref{g11}, \eqref{g12} defining respectively  local and nonlocal spectral conductances are presented with excess accuracy. This is because the boundary conditions \eqref{gplus}-\eqref{gminus} employed here remain applicable only in the tunneling limit and for weak spin dependent scattering $|G_m|, |G_{\varphi}| \ll G_T$. Hence, strictly speaking only the lowest order terms in $G_{m_{1,2}}$ and $G_{\varphi_{1,2}}$ need to be kept in our final results.

In order to proceed it is necessary to evaluate the interface conductances as well as the matrix functions $\hat{\mathcal{D}}^{T,L}_{1,2,S}$. Restricting ourselves to the second order in the interface transmissions we obtain
\begin{gather}
\hat g_{1T}(\varepsilon)=G_{T_1} \hat \nu_S(\bm{r}_1,\varepsilon)
+ G_{T_1}^2 \dfrac{\Delta^2\theta(\Delta^2-\varepsilon^2)}{\Delta^2-\varepsilon^2}
\hat U_1(\varepsilon),
\label{g1T}
\\
\hat g_{1L}(\varepsilon)=G_{T_1}\hat \nu_S(\bm{r}_1,\varepsilon)
-G_{T_1}^2 \dfrac{\Delta^2\theta(\varepsilon^2-\Delta^2)}{\varepsilon^2-\Delta^2}
\hat U_1(\varepsilon),
\label{g1L}
\\
\hat g_{1m}(\varepsilon)=G_{m_1} \hat \nu_S(\bm{r}_1,\varepsilon)\hat{\bm{\sigma}}\bm{m}_1,
\label{g1m}
\end{gather}
and analogous expressions for the interface conductances of the second interface. The matrix function
\begin{multline}
\hat U_1(\varepsilon)= \dfrac{D_1}{2\sigma_1}\Bigl\{ \Rea \left[
\mathcal{C}_1(\bm{r}_1,\bm{r}_1,2h_1^+) +
\mathcal{C}_1(\bm{r}_1,\bm{r}_1,2h_1^-) \right]
\\-
\hat{\bm{\sigma}}\bm{m}_1 \Rea \left[
\mathcal{C}_1(\bm{r}_1,\bm{r}_1,2h_1^+)-
\mathcal{C}_1(\bm{r}_1,\bm{r}_1,2h_1^-) \right]
\Bigr\}
\end{multline}
with $h_1^{\pm}=h_1 \pm \varepsilon$ defines the correction due to the proximity effect in the normal metal.

Taking into account the first order corrections in the interface transmissions one can derive the density of states inside the superconductor in the following form
\begin{multline}
\hat \nu_S(\bm{r},\varepsilon)=
\dfrac{|\varepsilon| \theta (\varepsilon^2 - \Delta^2)}{
\sqrt{|\varepsilon^2 - \Delta^2|}}
+\dfrac{D_S}{\sigma_S}\dfrac{\Delta^2}{\Delta^2 - \varepsilon^2}
\sum_{i=1,2}
\Biggl[
G_{T_i} \Rea \mathcal{C}_S(\bm{r},\bm{r}_i, 2\omega^R)
\\
-\hat{\bm{\sigma}}\bm{m}_i
G_{\varphi_i}\Img \mathcal{C}_S(\bm{r},\bm{r}_i, 2\omega^R)\Biggr],
\label{dos}
\end{multline}
where
\begin{equation}
\omega^R=
\begin{cases}
\sqrt{\varepsilon^2-\Delta^2}, \quad   &   \varepsilon>\Delta,
\\
i\sqrt{\Delta^2 -\varepsilon^2}, \quad &  |\varepsilon| < \Delta,
\\
-\sqrt{\varepsilon^2-\Delta^2}, \quad  &  \varepsilon < \Delta,
\end{cases}
\label{omegar}
\end{equation}
and the Cooperon $\mathcal{C}_{j}(\bm{r}, \bm{r}^{\prime}, \varepsilon)$ represents the solution of the equation
\begin{equation}
\left(-D\nabla^2-i\varepsilon\right)\mathcal{C}(\bm{r}, \bm{r}^{\prime}, \varepsilon)=
\delta(\bm{r}- \bm{r}^{\prime})
\end{equation}
in the normal metal leads ($j=1,2$) and the superconductor ($j=S$).
In the quasi-one-dimensional geometry the corresponding solutions take the form
\begin{gather}
\mathcal{C}_j (x_j, x_j, \varepsilon) =
\dfrac{\tanh\left(k_j L_j\right)}{S_jD_jk_j},
\quad j=1,2,
\\
\mathcal{C}_S (x,x',\varepsilon)=
\dfrac{\sinh [k_S(L-x')]\sinh k_S x}{k_S S_S D_S\sinh (k_S L)}, \quad x'>x,
\end{gather}
where $S_{S,1,2}$ are the wire cross sections and $k_{1,2,S}=\sqrt{-i\varepsilon/D_{1,2,S}}$.

Substituting Eq. (\ref{dos}) into Eqs. (\ref{g1T}) and (\ref{g1L}) and comparing the terms $\propto G_{T_1}^2$ we observe that the tunneling correction to the density of states dominates over the terms proportional to $\hat U_1$ which contain an extra small factor $\sqrt{\Delta/h} \ll 1$. Hence, the latter terms in Eqs. (\ref{g1T}) and (\ref{g1L}) can be safely neglected. In addition, in Eq. \eqref{dos} we also neglect small tunneling corrections to the superconducting density of states at energies exceeding the superconducting gap $\Delta$. Within this approximation the density of states inside the superconducting wire becomes spin-independent $\hat \nu_S (\bm{r} , \varepsilon) = \hat \sigma_0 \nu_S( \bm{r} , \varepsilon)$. It can then be written as
\begin{multline}
\nu_S(\bm{r},\varepsilon)=
\dfrac{|\varepsilon|}{\sqrt{|\varepsilon^2 - \Delta^2|}}
\theta(\varepsilon^2-\Delta^2)
\\
+\dfrac{D_S}{\sigma_S}\dfrac{\Delta^2\theta(\Delta^2-\varepsilon^2)}{\Delta^2 - \varepsilon^2}
\sum\limits_{i=1,2}
G_{T_i} \Rea \mathcal{C}_S(\bm{r},\bm{r}_i, 2\omega^R).
\label{dos1}
\end{multline}
Accordingly, the interface conductances take the form
\begin{gather}
\hat g_{1T}(\varepsilon) = \hat g_{1L}(\varepsilon)=G_{T_1}\nu_S(\bm{r}_1,\varepsilon),
\label{g1TL1}
\\
\hat g_{1m}(\varepsilon)=G_{m_1} \nu_S(\bm{r}_1,\varepsilon)\hat{\bm{\sigma}}\bm{m}_1.
\label{g1m1}
\end{gather}
Let us emphasize again that within our approximation the $G_{\varphi}$-term does not enter into expressions for the interface conductances \eqref{g1TL1}-\eqref{g1m1} and, hence, does not appear in the final expressions for the conductances $g_{ij}(\varepsilon)$.

In the limit of strong exchange fields $h_{1,2} \gg \Delta$ and small interface transmissions considered here the  proximity effect in the ferromagnets remains weak and can be neglected. Hence, the functions $\hat{\mathcal{D}}^{T,L}_{1}(\bm{r}_1,\bm{r}_1,\varepsilon)$ and $\hat{\mathcal{D}}^{T,L}_{2}(\bm{r}_2,\bm{r}_2,\varepsilon)$ can be approximated by their normal state values
\begin{gather}
\hat{\mathcal{D}}^{T,L}_{1}(\bm{r}_1,\bm{r}_1,\varepsilon)=\sigma_1 r_{N_1}\hat 1/D_1,
\\
\hat{\mathcal{D}}^{T,L}_{2}(\bm{r}_2,\bm{r}_2,\varepsilon)=\sigma_2 r_{N_2}\hat 1/D_2,
\\
r_{N_j} = L_j/(\sigma_j S_j), \quad j=1,2,
\end{gather}
where $r_{N_1}$ and $r_{N_2}$ are the normal state resistances of ferromagnetic terminals.
In the the superconducting region an effective expansion parameter is
$G_{T_{1,2}}r_{\xi_S}(\varepsilon)$, where
$r_{\xi_S}(\varepsilon)=\xi_S(\varepsilon)/(\sigma_S S_S)$ is the Drude
resistance of the superconducting wire segment of length $\xi_S(\varepsilon)=\sqrt{D_S/2|\omega^R|}$ and $\omega^R$ is the function of $\varepsilon$ according to Eq.~\eqref{omegar}. In the limit
\begin{equation}
G_{T_{1,2}}r_{\xi_S}(\varepsilon) \ll 1,
\end{equation}
which is typically well satisfied for realistic system parameters, it suffices to evaluate the function $\hat{\mathcal{D}}^T_S(x,x',\varepsilon)$ for impenetrable interfaces. In this case we find
\begin{equation}
\hat{\mathcal{D}}_S^T (x,x',\varepsilon)=
\begin{cases}
\dfrac{\Delta^2 - \varepsilon^2}{\Delta^2}\mathcal{C}_S (x,x',2\omega^R),  &|\varepsilon|  < \Delta,
\\
\dfrac{\varepsilon^2 -\Delta^2 }{\varepsilon^2}\mathcal{C}_S (x,x',0),  &|\varepsilon| > \Delta.
\end{cases}
\end{equation}
We note that special care should be taken while calculating $\mathcal{D}^L_S(x,x',\varepsilon)$ at subgap energies, since the coefficient $D^L$ in Eq. \eqref{diffeqL} tends to zero deep inside the superconductor. Accordingly, the function $\mathcal{D}^L_S(x,x',\varepsilon)$ becomes singular in this case.  Nevertheless, the combinations $\hat R_j^L(\mathcal{M}^L)^{-1}$ and $\hat R_{12}^L(\mathcal{M}^L)^{-1}$ remain finite also in this limit. At subgap energies we obtain
\begin{multline}
\hat R_1^L(\hat{\mathcal{M}}^L)^{-1}=
\hat R_2^L(\hat{\mathcal{M}}^L)^{-1}=
\hat R_{12}^L(\hat{\mathcal{M}}^L)^{-1}
\\=
\dfrac{1}{r_{N_1}+r_{N_2}+
\dfrac{2\kappa e^{d/\xi_S(\varepsilon)}}{
r_{\xi_S}(\varepsilon) G_{T_1} G_{T_2}}},
\end{multline}
where $\kappa =1-\varepsilon^2/\Delta^2$ and $d=|x_2-x_1|$ is the distance between two FS contacts.
Substituting the above relations into Eq.  \eqref{g12}
we arrive at the final result for the non-local spectral conductance of our device at subgap energies ($|\varepsilon|< \Delta$)
\begin{multline}
g_{12}(\varepsilon)=g_{21}(\varepsilon)=
\dfrac{\kappa r_{\xi_S}(\varepsilon) \exp[-d/\xi_S (\varepsilon)]}{
2[r_{N_1} + 1/g_{T1}(\varepsilon)][r_{N_2} + 1/g_{T2}(\varepsilon)]}
\\
\times\left[
1+\frac{\bm{m}_1\bm{m}_2}{\kappa}
\dfrac{G_{m1}}{g_{T1}(\varepsilon)}\dfrac{G_{m2}}{g_{T2}(\varepsilon)}
\dfrac{1}{\kappa + \dfrac{r_{N_1}+r_{N_2}}{2}
r_{\xi_S}(\varepsilon) G_{T_1} G_{T_2} e^{-d/\xi_S(\varepsilon)} }
\right].
\label{g12e}
\end{multline}

Eq. \eqref{g12e} represents the central result of this section. It consists of two different contributions. The first of them is independent of the interface polarizations $\bm{m}_{1,2}$. This term represents direct generalization of the result \cite{GKZ} in two different aspects.  Firstly, the analysis \cite{GKZ}
was carried out under the assumption $r_{N_{1,2}}g_{T1,2}(\varepsilon) \ll 1$ which is abandoned here. Secondly (and more importantly), sufficiently large exchange fields $h_{1,2} \gg \Delta$ of ferromagnetic electrodes suppress disorder-induced electron interference in these electrodes and, hence, eliminate the corresponding zero-bias anomaly both in local \cite{VZK,HN,Z} and non-local \cite{GKZ} spectral conductances. In this case with sufficient accuracy one can set $g_{Ti}(\varepsilon)=G_{Ti}\nu_S(x_i,\varepsilon)$ implying that at subgap energies
$g_{Ti}(\varepsilon)$ is entirely determined by the second term in Eq. (\ref{dos1})
which yields in the case of quasi-one-dimensional electrodes
\begin{gather}
g_{T1}(\varepsilon)=\dfrac{\Delta^2 G_{T_1}r_{\xi_S}(\varepsilon)}{2(\Delta^2-\varepsilon^2)}
\left[ G_{T_1} + G_{T_2}e^{-d/\xi_S (\varepsilon)} \right],
\\
g_{T2}(\varepsilon)=\dfrac{\Delta^2 G_{T_2}r_{\xi_S}(\varepsilon)}{2(\Delta^2-\varepsilon^2)}
\left[ G_{T_2} + G_{T_1}e^{-d/\xi_S (\varepsilon)} \right].
\end{gather}

Note, that
if the exchange field $h_{1,2}$ in both normal electrodes is reduced well below $\Delta$ and eventually is set equal to zero, the term containing $\hat U_1(\varepsilon )$ in Eqs. (\ref{g1T}), (\ref{g1L}) becomes important and should be taken into account. In this case we again recover the zero-bias anomaly \cite{VZK,HN,Z} $g_{Ti}(\varepsilon) \propto 1/\sqrt{\varepsilon}$ and from the first term in Eq. \eqref{g12e} we reproduce the results \cite{GKZ} derived in the limit $h_{1,2} \to 0$.

\begin{figure}
\centerline{\includegraphics[width=7.5cm]{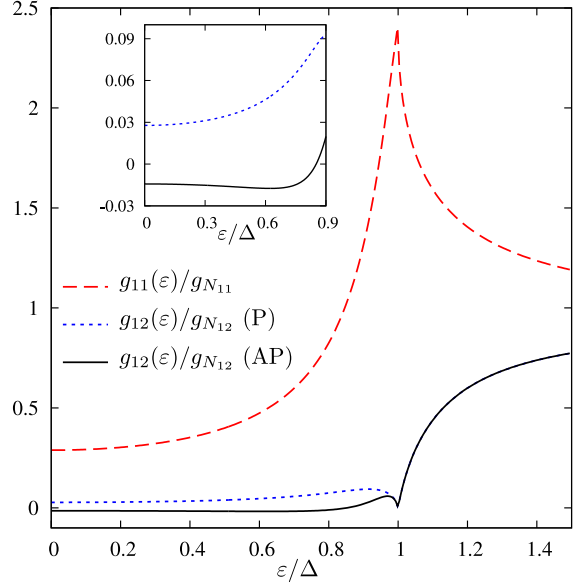}}
\caption{Local (long-dashed line) and non-local (short-dashed and solid lines) spectral conductances normalized to its normal state values. Here we choose $r_{N_1} = r_{N_2} = 5 r_{\xi_S}(0)$, $x_1 = L-x_2 = 5 \xi_S(0)$, $x_2-x_1=\xi_S (0)$, $G_{T_1} = G_{T_2} = 4G_{m_1} = 4G_{m_2}=0.2/r_{\xi_S}(0)$. Energy dependence of non-local conductance is displayed for parallel (P) $\bm{m}_1\bm{m}_2=1$ and antiparallel (AP) $\bm{m}_1\bm{m}_2=-1$ interface magnetizations. Inset: The same in the limit of low energies.}
\label{fig-e}
\end{figure}

The second term in \eqref{g12e} is proportional to the product $\bm{m}_1\bm{m}_2 G_{m1} G_{m2}$ and
describes non-local magnetoconductance effect in our system emerging due to spin-sensitive electron scattering at FS interfaces. It is important that -- despite the strong inequality $|G_{mi}| \ll G_{Ti}$ -- both terms in Eq. \eqref{g12e} can be of the same order, i.e. the second (magnetic) contribution can significantly modify the non-local conductance of our device.

In the limit of large interface resistances $r_{N_{1,2}}g_{T1,2}(\varepsilon) \ll 1$ the formula \eqref{g12e} reduces to a much simpler one
\begin{multline}
g_{12}(\varepsilon)=g_{21}(\varepsilon)=
\dfrac{r_{\xi_S}(\varepsilon)}{2}
\exp[-d/\xi_S (\varepsilon)]
\label{g12e1}\\
\times\left[
\dfrac{\Delta^2 - \varepsilon^2}{\Delta^2}
g_{T1}(\varepsilon)g_{T2}(\varepsilon)+
\bm{m}_1\bm{m}_2 G_{m1} G_{m2}
\dfrac{\Delta^2}{\Delta^2-\varepsilon^2}\right].
\end{multline}
Interestingly, Eq. (\ref{g12e1}) remains applicable for arbitrary values of the angle between interface polarizations $\bm{m}_1$ and $\bm{m}_2$ and strongly resembles the analogous result for the non-local conductance in ballistic FSF systems (cf., e.g., Eq. (\ref{G12zeroVzeroTheta}) in the previous section). The first term in the square brackets in Eq. \eqref{g12e1} describes the fourth order contribution in the interface transmissions which remains nonzero also in the limit of the nonferromagnetic leads \cite{GKZ}. In contrast, the second term is proportional to the product of transmissions of both interfaces, i.e. only to the second order in barrier transmissions. This term vanishes identically provided at least one of the interfaces is spin-isotropic.

Contrary to the non-local conductance at subgap energies, both local conductance (at all energies) and non-local spectral conductance at energies above the superconducting gap are only weakly affected by magnetic effects. Neglecting small corrections due to $G_m$ term in the boundary conditions we obtain
\begin{gather}
\hat g_{11}(\varepsilon)=\hat R_1^T(\hat{\mathcal{M}}^T)^{-1}, \quad
\hat g_{22}(\varepsilon)=\hat R_2^T(\hat{\mathcal{M}}^T)^{-1},
\label{g11c}
\\
\hat g_{12}(\varepsilon)=g_{21}(\varepsilon)=\hat R_{12}^T(\hat{\mathcal{M}}^T)^{-1}, \quad
|\varepsilon| > \Delta.
\label{g12c}
\end{gather}

Eqs. \eqref{g11c} and \eqref{g12c} together with the above expressions for the  non-local subgap conductance enable one to recover both local and non-local spectral conductances of our system at all energies. Typical energy dependencies for both $g_{11}(\varepsilon)$ and $g_{12}(\varepsilon)$ are displayed in Fig.~\ref{fig-e}.
For instance, we observe that at subgap energies the non-local conductance $g_{12}$ changes its sign being positive for parallel and negative for antiparallel interface polarizations.

\begin{figure}
\centerline{\includegraphics[width=7.5cm]{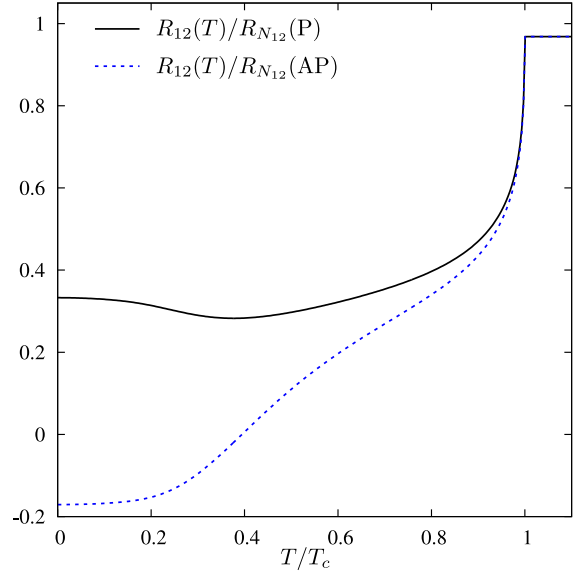}}
\caption{Non-local resistance (normalized to its normal state value) versus temperature (normalized to the superconducting critical temperature $T_C$) for parallel (P) and antiparallel (AP) interface magnetizations. The parameters are the same as in Fig.~\ref{fig-e}.}
\label{fig-r}
\end{figure}

\begin{figure}
\centerline{\includegraphics[width=7.5cm]{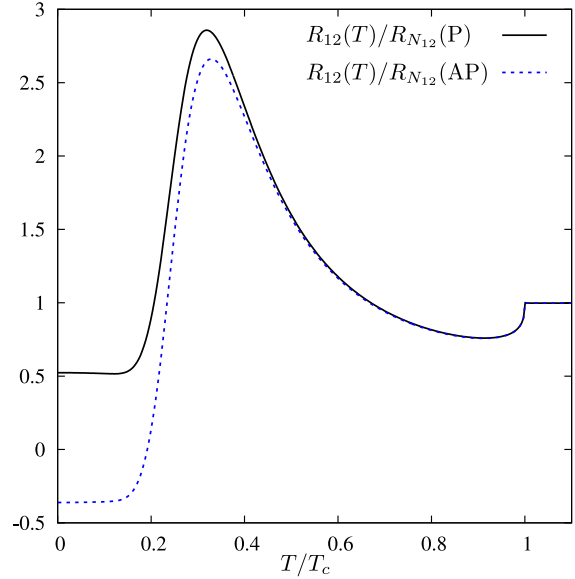}}
\caption{The same as in Fig.~\ref{fig-r} for the following parameter values: $r_{N_1} = r_{N_2} = 5 r_{\xi_S}(0)$, $x_1 = L-x_2 = 5 \xi_S(0)$, $x_2-x_1=\xi_S (0)$, $G_{T_1} = G_{T_2} = 25G_{m_1} = 25G_{m_2}=0.025/r_{\xi_S}(0)$. }
\label{fig-r1}
\end{figure}

\subsection{I-V curves}

Having established the spectral conductance matrix $g_{ij}(\varepsilon)$ one can easily recover the complete $I-V$ curves for our hybrid FSF structure. In the limit of low bias voltages these $I-V$ characteristics become linear, i.e.
\begin{gather}
I_1 = G_{11}(T) V_1 - G_{12}(T) V_2,
\label{I1}
\\
I_2 = - G_{21}(T) V_1 + G_{22}(T) V_2,
\label{I2}
\end{gather}
where $G_{ij}(T)$ represent the linear conductance matrix defined as
\begin{equation}
G_{ij}(T)=\dfrac{1}{4T}
\int g_{ij}(\varepsilon)\dfrac{d\varepsilon}{\cosh^2\dfrac{\varepsilon}{2T}}.
\end{equation}
It may also be convenient to invert the relations \eqref{I1}-\eqref{I2} thus expressing induced voltages $V_{1,2}$ in terms of injected currents $I_{1,2}$:
\begin{gather}
V_1 = R_{11}(T) I_1 + R_{12}(T) I_2,
\label{V1}
\\
V_2 = R_{21}(T) I_1 + R_{22}(T) I_2,
\label{V2}
\end{gather}
where the coefficients $R_{ij}(T)$ define local ($i=j$) and nonlocal ($i\neq j$) resistances
\begin{gather}
R_{11}(T)=\dfrac{G_{22}(T)}{G_{11}(T)G_{22}(T)-G_{12}^2(T)},
\\
R_{12}(T)=R_{21}(T)=\dfrac{G_{12}(T)}{G_{11}(T)G_{22}(T)-G_{12}^2(T)}
\end{gather}
and similarly for $R_{22}(T)$. In non-ferromagnetic NSN structures the low temperature non-local resistance  $R_{12}(T\to 0)$ turns out to be independent of both the interface conductances and the parameters of the normal leads \cite{GKZ}. However, this universality of $R_{12}$ does not hold anymore provided
non-magnetic normal metal leads are substituted by ferromagnets.  Non-local linear resistance $R_{12}$ of our FSF structure is displayed in Figs.~\ref{fig-r}, \ref{fig-r1} as a function of temperature for parallel ($\bm{m}_1\bm{m}_2=1$) and antiparallel ($\bm{m}_1\bm{m}_2=-1$) interface magnetizations. In Fig.~\ref{fig-r} we show typical temperature behavior of the non-local resistance for sufficiently transparent interfaces. For both mutual interface magnetizations $R_{12}$ first decreases with temperature below $T_C$ similarly to the non-magnetic case. However, at lower $T$ important differences occur: While in the case of parallel magnetizations $R_{12}$ always remains positive and even shows a noticeable upturn at sufficiently low $T$, the non-local resistance for antiparallel magnetizations keeps monotonously decreasing with $T$ and may become negative in the low temperature limit. In the limit of very low interface transmissions the temperature dependence of the non-local resistance exhibits a well pronounced charge imbalance peak (see Fig.~\ref{fig-r1}) which physics is similar to that analyzed in the case of non-ferromagnetic NSN structures \cite{GZ07,KZ08,GKZ}. Let us point out that the above behavior of the non-local resistance is qualitatively consistent with available experimental observations \cite{Beckmann}. More experiments would be desirable in order to quantitatively verify our theoretical predictions.

\section{Concluding remarks}

In this paper we developed a non-perturbative theory of non-local
electron transport in both ballistic and diffusive NSN and FSF
three-terminal structures with spin-active interfaces. Our theory
is based on the quasiclassical formalism of energy-integrated
Green-Eilenberger functions supplemented by appropriate
boundary conditions describing spin-dependent scattering at
NS and FS interfaces. Our approach applies
at arbitrary interface transmissions and allows to fully describe
non-trivial interplay between spin-sensitive normal scattering,
local and non-local Andreev reflection at NS and FS interfaces.

In the case of ballistic structures our main
results are the general expressions for the non-local
cross-current $I_{12}$, Eq. \eqref{I12phi}, and for the non-local
correction $I_{11}$ to the BTK current, Eq. \eqref{CARcorr}. These
expressions provide complete description of the conductance matrix
of our three-terminal NSN device at arbitrary voltages,
temperature, spin-dependent transmissions of NS interfaces and
their polarizations. One of our important observations
is that in the case of ballistic electrodes no crossed Andreev
reflection can occur in both NSN and HSH structures with fully
open interfaces. Beyond the tunneling limit the dependence of the
non-local conductance on the size of the S-electrode $L$ is in
general non-exponential and reduces to $G_{12} \propto \exp
(-2L\Delta /v_F )$ only in the limit of large $L$. For hybrid
structures half-metal-superconductor-half-metal we predict that
the low energy non-local conductance does not vanish already in
the lowest order in barrier transmissions $G_{12} \propto
D_{1\uparrow}D_{2\uparrow}$.

In the second part of our paper we addressed spin-resolved non-local
electron transport in FSF structures in the presence
of disorder in the electrodes. Within our model transfer of electrons
across FS interfaces is described in the tunneling limit and
magnetic properties of the system
are accounted for by introducing ($i$) exchange fields $\bm{h}_{1,2}$
in both normal metal electrodes and ($ii$) magnetizations  $\bm{m}_{1,2}$
of both FS interfaces. The two
ingredients ($i$) and ($ii$) of our model
are in general independent from each other and have different physical
implications. While the role of (comparatively large) exchange fields
$h_{1,2}\gg \Delta$ is merely to suppress disorder-induced interference
of electrons \cite{VZK,HN,Z} penetrating from a superconductor into
ferromagnetic electrodes, spin-sensitive electron scattering at FS
interfaces yields an extra contribution to the non-local conductance
which essentially depends on relative orientations of the interface
magnetizations. For anti-parallel magnetizations the total non-local
conductance $g_{12}$ and resistance $R_{12}$ can turn negative at
sufficiently low energies/temperatures. At higher temperatures the
difference between the values of $R_{12}$ evaluated for
parallel and anti-parallel magnetizations becomes less important. At
such temperatures the non-local resistance behaves similarly to the
non-magnetic case demonstrating, e.g., a well-pronounced charge
imbalance peak \cite{GKZ} in the limit of low interface transmissions.

Our predictions can be directly used for quantitative
analysis of experiments on non-local electron transport in
hybrid FSF structures.

\vspace{0.5cm}

\centerline{\bf Acknowledgments}

\vspace{0.5cm}

This work was supported in part by DFG and by RFBR grant 09-02-00886. M.S.K. also acknowledges support from the Council for grants of the Russian President (Grant No. 89.2009.2) and from the Dynasty Foundation.

\end{document}